\documentclass[apj]{emulateapj}
\usepackage[utf8]{inputenc}
\usepackage{lipsum}
\usepackage{pifont}
\usepackage{amsmath, amssymb, bm}
\usepackage{balance}
\usepackage{multirow}
\usepackage{color}
\usepackage{enumitem}

\usepackage[usenames,dvipsnames]{xcolor}
\usepackage{hyperref}
\hypersetup{
    colorlinks = true,
    citecolor = {MidnightBlue},
    linkcolor = {BrickRed},
    urlcolor = {black}		
}

\slugcomment{Draft version}

\def\be{\begin{equation}}
\def\ee{\end{equation}}
\def\bea{\begin{eqnarray}}
\def\eea{\end{eqnarray}}

\begin{document}

\title{Weighing neutrinos with cosmic neutral hydrogen}

\author{Francisco Villaescusa-Navarro}
\affiliation{INAF - Osservatorio Astronomico di Trieste, Via G.B. Tiepolo 11, I-34143 Trieste, Italy\\
INFN, Sezione di Trieste, Via Valerio 2, I-34127 Trieste, Italy}
\email{villaescusa@oats.inaf.it}
\author{Philip Bull}
\affiliation{Institute of Theoretical Astrophysics, University of Oslo, PO Boks 1029 Blindern, 0315 Oslo, Norway}
\email{p.j.bull@astro.uio.no}
\author{Matteo Viel}
\affiliation{INAF - Osservatorio Astronomico di Trieste, Via G.B. Tiepolo 11, I-34143 Trieste, Italy\\
INFN, Sezione di Trieste, Via Valerio 2, I-34127 Trieste, Italy}
\email{viel@oats.inaf.it}

\date{\today}

\begin{abstract}

We investigate the signatures left by massive neutrinos on the spatial distribution of neutral hydrogen (HI) in the post-reionization era by running hydrodynamic simulations that include massive neutrinos as additional collisionless particles. We find that halos in massive/massless neutrino cosmologies host a similar amount of neutral hydrogen, although for a fixed halo mass, on average, the HI mass increases with the sum of the neutrino masses. Our results show that HI is more strongly clustered in cosmologies with massive neutrinos, while its abundance, $\Omega_{\rm HI}(z)$, is lower. These effects arise mainly from the impact of massive neutrinos on cosmology: they suppress both the amplitude of the matter power spectrum on small scales and the abundance of dark matter halos. Modelling the HI distribution with hydrodynamic simulations at $z > 3$, and a simple analytic model at $z<3$, we use the Fisher matrix formalism to conservatively forecast the constraints that Phase 1 of the Square Kilometre Array (SKA) will place on the sum of neutrino masses, $M_\nu\equiv \Sigma m_{\nu}$. We find that with 10,000 hours of interferometric observations at $3 \lesssim z \lesssim 6$ from a deep and narrow survey with SKA1-LOW, the sum of the neutrino masses can be measured with an error $\sigma(M_\nu)\lesssim0.3$ eV (95\% CL). Similar constraints can be obtained with a wide and deep SKA1-MID survey at $z \lesssim 3$, using the single-dish mode. By combining data from MID, LOW, and Planck, plus priors on cosmological parameters from a Stage IV spectroscopic galaxy survey, the sum of the neutrino masses can be determined with an error $\sigma(M_\nu)\simeq0.06$ eV (95\% CL).

\end{abstract}

\keywords{massive neutrinos, intensity mapping, cosmology} 

\maketitle

\section{Introduction}
\label{sec:introduction}

The standard model of particle physics describes neutrinos as neutral spin-$1/2$, massless fermions, organized into three families with three different flavours: $\nu_e, \nu_\mu, \nu_\tau$. It has been observed that neutrinos can change their flavour as they propagate through space, however. This phenomenon, known as neutrino oscillations, implies that neutrinos are massive. Measurements of the neutrino oscillations from laboratory experiments have 
allowed us to estimate the mass-square differences among the different neutrino mass eigenstates to be \citep{Fogli,Tortola}:
\bea
\bigtriangleup m^2_{12}&=&7.5\times10^{-5} ~{\rm eV}^2\\
|\bigtriangleup m^2_{23}|&=&2.3\times10^{-3}~{\rm eV}^2,
\eea
which implies that at least two of the three neutrino families are massive. A lower bound on the sum of the neutrino masses can be set from the above measurements: $M_\nu\equiv\sum_i m_{\nu_i}\gtrsim0.06$ eV. Unfortunately, the above constraints do not allow us to determine which neutrino is the lightest, or whether it is massless or massive. This gives rise to two different \textit{hierarchies}: a normal hierarchy in which $0\leqslant m_1\textless m_2\textless m_3$, and an inverted hierarchy where $0\leqslant m_3\textless m_1\textless m_2$. 

The fact that neutrinos are massive is one of the clearest indications of physics beyond the particle physics standard model, and so two of the most important questions in modern physics are: (a) what are the masses of the neutrinos, and (b) which hierarchy do they conform to? Answering these questions with laboratory experiments is extremely challenging. For instance, current bounds on the mass of the electron anti-neutrino from the KATRIN\footnote{\url{https://www.katrin.kit.edu/}} experiment are $m(\bar{\nu}_e)\textless2.3$ eV \citep{Kraus_2005} and are expected to improve to $m(\bar{\nu}_e)\textless0.2$ eV in the coming years.

On the other hand, tight upper limits on the neutrino masses have already been obtained by using cosmological observables such as the anisotropies in the cosmic microwave background (CMB), the clustering of galaxies, the abundance of galaxy clusters, the distortion in the shape of galaxies by weak lensing, the Ly$\alpha$ forest, etc. \citep{Hannestad_2003,Reid,Thomas,Swanson,Saito_2010,2011APh....35..177A,dePutter, Xia2012,WiggleZ,Zhao2012,
Costanzi,Basse,Planck_2013, Costanzi_2014, Costanzi_2014b, Wyman_2013,Battye_2013,Hamann_2013, Beutler_2014,Giusarma_2014, Palanque-Delabrouille:2014jca,Planck_2015, ades}. These constraints arise because massive neutrinos delay the matter-radiation equality time and slow down the growth of matter perturbation on small scales. At linear order, the resulting effects on the CMB and matter power spectrum are well known and understood, making cosmological observables extremely useful tools for putting upper limits on the sum of the neutrino masses.

Currently, the tightest upper limit on the sum of the neutrino masses comes from combining data from the CMB, baryonic acoustic oscillations (BAO) and the Ly$\alpha$ forest: $M_\nu<0.12$ eV ($95\%$ CL) \citep{ades}. These limits also have strong implications for particle physics experiments like neutrinoless double beta decay \citep{delloro}.

A new cosmological observable has recently been proposed that is expected to play an important role in future cosmology: 21cm intensity mapping \citep{Bharadwaj_2001A, Bharadwaj_2001B, Battye:2004re,McQuinn_2006, Chang_2008,Loeb_Wyithe_2008, Bull_2015}. The idea of this technique is to measure the integrated 21cm emission from unresolved galaxies by performing a low angular resolution survey \citep{Santos_2015}.  Since neutral hydrogen (HI) is a tracer of the underlying matter distribution of the Universe on large scales, the HI power spectrum is expected  to follow the shape of the matter power spectrum, but with a different amplitude (the HI bias). This should allow tight constraints to be placed on cosmological parameters through measurements of the power spectrum of the 21cm field \citep{Bull_2015}.

Intensity mapping therefore constitutes a promising new cosmological observable that can be used to constrain the neutrino masses \citep{Loeb_Wyithe_2008,2008PhRvD..78f5009P, Tegmark_2008, Metcalf_2010, 2011APh....35..177A,  Oyama_2012,Shimabukuro_2014}. In order to do that, however, one needs to understand how the 21cm power spectrum is affected by the presence of massive neutrinos. The aim of this paper is to investigate the signatures left by massive neutrinos on the 21cm power spectrum in the post-reionization universe, in both the linear and fully non-linear regimes.

We begin by studying the effects that massive neutrinos have on the spatial distribution of neutral hydrogen in real-space. We do this by running hydrodynamic simulations with massless and massive neutrinos. We investigate how the presence of massive neutrinos affects the HI abundance and clustering properties and, ultimately, the signatures left by neutrinos in the 21cm power spectrum.

We also forecast the constraints that the future Square Kilometre Array (SKA) radio telescope will place on the sum of the neutrino masses. We do this using the Fisher matrix formalism, where the spatial distribution of neutral hydrogen is modeled using hydrodynamic simulations at redshifts $3\leqslant z \leqslant 5.5$ (a redshift range where SKA1-LOW will collect data), and with a simple analytic model at redshifts $z\leqslant3$ (the redshift range covered by SKA1-MID).  Our approach is conservative in the sense that: (i) at $z>3$ we compare models of the HI distribution using 4 different methods; and (ii) we embed the information from the 21cm power spectrum in the Fisher matrix forecasts in a conservative way.

This paper is organized as follows. In Sec. \ref{sec:N-body} we describe the set of hydrodynamic simulations carried out for this work. Our simulations do not account for two crucial processes needed to properly model the spatial distribution of neutral hydrogen: HI self-shielding, and the formation of molecular hydrogen. We correct the outputs of our simulations {\it a posteriori} to account for these effects, depicting the four different methods we use to achieve this in Sec. \ref{sec:HI}. We also describe the method we use to model the spatial distribution of neutral hydrogen at redshifts $z\leqslant3$, which is not covered by our hydrodynamic simulations. In Sec. \ref{sec:neutrinos_effects} we investigate the effect of massive neutrinos on the abundance and spatial distribution of neutral hydrogen. We present our forecasts on the neutrino masses in Sec. \ref{sec:forecasts} and, finally, draw the main conclusions of this paper in Sec. \ref{sec:conclusions}.

\section{Hydrodynamic simulations} 
\label{sec:N-body}

\begin{table*}[t]
\begin{center}
{\renewcommand{\arraystretch}{1.5}
\resizebox{16cm}{!}{
\begin{tabular}{|c|c|c|c|c|c|c|c|c|c|c|}
\hline
Name & Box & $\Omega_{\rm cdm}$ & $\Omega_{\rm b}$ & $\Omega_\nu$ & $\Omega_\Lambda$ & $\Omega_k$ & $h$ & $n_s$ & $10^{9}A_s$ & $\sigma_{8,0}$\\
 & ($h^{-1}\rm{Mpc}$) & & & & & & & & & \\
\hline
\hline 
$\mathcal{F}$ & 50 & $0.2685$ & $0.049$ & $0.0$ & $0.6825$ & 0 & $0.67$ & $0.9624$ & $2.13$ & $0.834$\\
\hline
$\mathcal{\nu}^{+}$ & 50 & $0.2685$ & $0.049$ & $0.007075$ & $0.675425$ &  0 & $0.67$ & $0.9624$ & $2.13$  & $0.778$\\ 
$\mathcal{\nu}_{\rm m}^{+}$ & 50 & $0.261425$ & $0.049$ & $0.007075$ & $0.6825$ &  0 & $0.67$ & $0.9624$ & $2.13$  & $0.764$\\ 
$\mathcal{\nu}_{\rm m}^{++}$ & 50 & $0.25435$ & $0.049$ & $0.01415$ & $0.6825$ & 0 & $0.67$ & $0.9624$ & $2.13$  & $0.693$\\ 
\hline
$\mathcal{C}^{+}$ & 50 & $0.287$ & $0.049$ & $0.0$ & $0.664$ & 0 & $0.67$ & $0.9624$ & $2.13$  & $0.868$\\ 
$\mathcal{C}^{-}$ & 50 & $0.25$ & $0.049$ & $0.0$ & $0.701$ & 0 & $0.67$ & $0.9624$ & $2.13$  & $0.797$\\ 
\hline
$\mathcal{B}^{+}$ & 50 & $0.2685$ & $0.055$ & $0.0$ & $0.6765$ & 0 & $0.67$ & $0.9624$ & $2.13$  & $0.816$\\ 
$\mathcal{B}^{-}$ & 50 & $0.2685$ & $0.043$ & $0.0$ & $0.6885$ & 0 & $0.67$ & $0.9624$ & $2.13$  & $0.853$\\ 
\hline
$\mathcal{H}^{+}$ & 50 & $0.2685$ & $0.049$ & $0.0$ & $0.6825$ & 0 & $0.71$ & $0.9624$ & $2.13$  & $0.886$\\
$\mathcal{H}^-$ & 50 & $0.2685$ & $0.049$ & $0.0$ & $0.6825$ & 0 & $0.63$ & $0.9624$ & $2.13$  & $0.777$\\
\hline
$\mathcal{N}^+$ & 50 & $0.2685$ & $0.049$ & $0.0$ & $0.6825$ & 0 & $0.67$ & $1.0009$ & $2.13$ & $0.846$\\
$\mathcal{N}^-$ & 50 & $0.2685$ & $0.049$ & $0.0$ & $0.6825$ & 0 & $0.67$ & $0.9239$ & $2.13$ & $0.822$\\
\hline
$\mathcal{A}^+$ & 50 & $0.2685$ & $0.049$ & $0.0$ & $0.6825$ & 0 & $0.67$ & $0.9624$ & $2.45$ & $0.894$\\
$\mathcal{A}^-$ & 50 & $0.2685$ & $0.049$ & $0.0$ & $0.6825$ & 0 & $0.67$ & $0.9624$ & $1.81$ & $0.769$\\
\hline
\end{tabular}
}}
\end{center} 
\caption{Summary of our simulation suite. The simulation name indicates the parameter that has been varied with respect to the fiducial, $\mathcal{F}$, model. The superscript, $^+/^-$, designates whether the variation is positive/negative. The simulations $\nu_{\rm m}^+$ and $\nu_{\rm m}^{++}$ are simulations with massive neutrinos having a value of $\Omega_{\rm m}$ equal to the one of the fiducial model.}
\label{tab_sims}
\end{table*}

We model the spatial distribution of neutral hydrogen by running high-resolution hydrodynamic simulations in 14 different cosmological models. The values of the cosmological parameters of our fiducial model are $\Omega_{\rm m}=0.3175$, $\Omega_{\rm cdm}=0.2865$, $\Omega_{\rm b}=0.049$, $\Omega_\nu=0$, $\Omega_\Lambda=0.6825$, $h=0.6711$, $n_s=0.9624$ and $\sigma_8=0.834$, in excellent agreement with the latest results from Planck \citep{Planck_2015}. In all simulations we have assumed a flat cosmology, and therefore the value of $\Omega_\Lambda$ is set to $1-\Omega_{\rm m}$, with $\Omega_{\rm m}=\Omega_{\rm cdm}+\Omega_{\rm b}+\Omega_\nu$, with $\Omega_\nu h^2\cong M_\nu/(94.1~{\rm eV})$. A summary of our simulation suite is shown in Table \ref{tab_sims}.

The simulations can be split into two different groups. On one hand we have simulations in which the value of one of the parameters, $\Omega_{\rm cdm}$, $\Omega_{\rm b}$, $\Omega_\nu$, $h$, $n_s$, $A_s$, is varied (with respect to the value in the fiducial model) while the values of the other parameters are kept fixed. We use these simulations in our Fisher matrix analysis to investigate degeneracies between cosmological parameters, and to forecast the constraints that the SKA will place on the neutrino masses. The simulations belonging to this group are $(\mathcal{F}, \nu^+, \mathcal{C}^+, \mathcal{C}^-, \mathcal{B}^+, \mathcal{B}^-, \mathcal{H}^+, \mathcal{H}^-, \mathcal{A}^+, \mathcal{A}^-)$.

On the other hand we have simulations in which we vary the value of $\Omega_\nu$, but keep $\Omega_{\rm m}$ fixed. This is the most natural choice to investigate the effect of massive neutrinos, since we assume that a fraction of the total matter content of the Universe is made up of neutrinos. We have run one simulation with $M_\nu=0.3$ eV, and another with $M_\nu=0.6$ eV. Even though these neutrino masses are ruled out by the most recent constraints that combine CMB, BAO and Ly$\alpha$-forest data, our purpose in this paper is to investigate the impact of neutrino masses on the HI spatial distribution. Note that for a realistic sum of neutrino masses, the effect will be very small and may be completely hidden by sample variance in our simulations. Therefore, we decided to run simulations with neutrino masses higher than current bounds to properly resolve the effects of neutrinos on the HI distribution. For the sum of the neutrino masses we use to run the simulations, 0.3 eV and 0.6 eV, the neutrino masses are almost perfectly degenerate, so there is no need to distinguish between the three different families. We use these simulations to investigate the impact of massive neutrinos on the matter and HI spatial distribution. The simulations belonging to this group are $(\mathcal{F},\nu_{\rm m}^+, \nu_{\rm m}^{++})$.

In each simulation we follow the evolution of $512^3$ CDM and $512^3$ baryon particles (plus $512^3$ neutrino particles for simulations with $\Omega_\nu>0$) in a periodic box of size $50$ comoving $h^{-1}$Mpc, down to redshift 3. For each simulation we save snapshots at redshifts 5.5, 5, 4.5, 4, 3.5 and 3. Note that evolving our simulations down to $z=0$, for all the cosmological models considered in this paper, would be extremely computationally expensive, so at redshifts lower than $z=3$ we model the spatial distribution of neutral hydrogen using a simple analytic model described in Sec. \ref{subsec:low-z}. 

The simulations were run using the TreePM+SPH code {\sc GADGET-III} \citep{Springel_2005}. They incorporate radiative cooling by hydrogen and helium, as well as heating by a uniform UV background. Both the cooling routine and the UV background have been modified to obtain the desired thermal history, which corresponds to
the reference model of \cite{viel13} that has been shown to provide a good fit to the statistical properties of the transmitted Lyman-$\alpha$ flux. In our simulations, hydrogen reionization takes place\footnote{Note that this reionization redshift is in slight tension with the latest Planck results \citep{Planck_2015, Mitra_2015}. We do not expect our conclusions to be affected by this however, since we use the same reionization history for all models, and are only interested in studying relative effects.} at $z\sim12$ and the temperature-density relation for the low-density IGM $T = T_0(z)(1 + \delta)^{
\gamma(z)-1}$ has $\gamma(z) = 1.3$ and $T_0(z = 2.4, 3, 4) = (16500, 15000, 10000)$ K. For every simulation we generate 5000 quasar mock spectra at the snapshot redshifts, and tune the strength of the UV background to reproduce the observed mean transmitted flux of the Lyman-$\alpha$ forest; this information is needed for two of the HI modeling methods (the pseudo-RT 1 and pseudo-RT 2 methods, described below). Star formation is modeled using the multi-phase effective model of \cite{Springel-Hernquist_2003}.

The simulation initial conditions are generated at $z=99$ using the Zel'dovich approximation. We compute the transfer functions of the different components using CAMB \citep{CAMB}. In simulations with massive neutrinos, the initial conditions were generated taking into account the scale-dependent growth present in those cosmological models. We note that the random seeds used to generate the initial conditions are the same in all simulations.

We identify dark matter halos using both the Friends-of-Friends (FoF) algorithm \citep{FoF} with $b=0.2$ and {\sc SUBFIND} \citep{Subfind,Dolag_2009}. We require that a minimum of 32 CDM particles belong to the FoF halo to identify it.

\begin{table*}
\centering{
{\renewcommand{\arraystretch}{1.8}
 \begin{tabular}{|l|cccc|}
 \hline
 & \multicolumn{4}{c|}{\bf{Method}} \\
 & Pseudo-RT 1 & Pseudo-RT 2 & Halo-based 1 & Halo-based 2 \\
 & (fiducial method) & & &\\
 \hline
 HI self-shielding & \large{\ding{52}} & \large{\ding{52}} & \large{\ding{52}} & \large{\ding{52}}\\
 Molecular hydrogen & \large{\ding{52}} & \large{\ding{52}} & \large{\ding{56}} & \large{\ding{56}}\\
  Modeling $M_{\rm HI}(M,z)$ function & \large{\ding{56}} & \large{\ding{56}} & \large{\ding{52}} & \large{\ding{52}}\\
HI assigned to all gas particles & \large{\ding{52}} & \large{\ding{52}} & \large{\ding{56}} & \large{\ding{56}}\\
HI assigned only to gas particles in halos & \large{\ding{56}} & \large{\ding{56}} & \large{\ding{52}} & \large{\ding{52}}\\
Value of $\Omega_{\rm HI}(z)$ fixed a priori & \large{\ding{56}} & \large{\ding{56}} & \large{\ding{52}} & \large{\ding{52}}\\
  Reference work & \cite{Rahmati_2013} & \cite{Dave_2013} & \cite{Bagla_2010} & This work\\
 \hline
 \end{tabular} }}
 \caption{Differences between the four different methods used to model the spatial distribution of neutral hydrogen.}
 \label{tbl:HI_methods}
\end{table*}

\section{HI distribution} 
\label{sec:HI}

Our hydrodynamic simulations do not take into account two crucial physical processes needed to properly simulate the spatial distribution of neutral hydrogen: the formation of molecular hydrogen (H$_2$), and HI self-shielding. In this section we describe the various methods we use to correct for those two processes. We consider four methods here: two models (pseudo-RT 1 and pseudo-RT 2) that aim to mimic the result of a full radiative transfer calculation\footnote{In both pseudo-RT methods, we only account for the radiation from the UV background and do not consider radiation from local sources \citep{Miralda-Escude_2005, Schaye_2006, Rahmati_sources}. We are interested only in studying relative differences here, rather than absolute quantities, and do not expect our conclusions to change by neglecting the radiation from local sources.}, and two (halo-based 1 and halo-based 2) that were constructed by taking into account the fact that all HI should be within dark matter halos \citep{Villaescusa-Navarro_2014a}. These models have the objective of providing the shape and amplitude of the function $M_{\rm HI}(M,z)$ (see subsection \ref{subsec:low-z} for further details). We summarize the main features of each method in Table \ref{tbl:HI_methods}.

In this section, we also explain the way we model the spatial distribution of neutral hydrogen at redshifts $z\textless3$. This redshift range is not covered by our simulations, but is needed to forecasts the constraints that the SKA will set on the neutrino masses.

\subsection{Pseudo-RT 1 (fiducial model)}

In the first \textit{pseudo-radiative transfer} method, neutral hydrogen is assigned to every single particle in the simulation. The HI self-shielding correction is modeled using the fitting formula of \cite{Rahmati_2013} (see their appendix A) that was obtained by performing radiative transfer calculations on top of hydrodynamic simulations. This correction states that the photo-ionization rate seen by a particular gas particle is a function of its density. 

The HI masses obtained in this way are further corrected to account for the presence of molecular hydrogen, which is only assigned to star-forming particles. We assume that the ratio of molecular to neutral hydrogen scales with the pressure, $P$, as
\begin{equation}
\frac{\Sigma_{H_2}}{\Sigma_{\rm HI}}=\left(\frac{P}{P_0}\right)^{\alpha}~,
\label{Things_Rmol}
\end{equation}
where $\Sigma_{\rm H_2}$ and $\Sigma_{\rm HI}$ are the molecular and neutral hydrogen surface densities, respectively. In our analysis we use $(P_0,\alpha)=(3.5\times10^4~{\rm cm^{-3}K}, 0.5)$, where $\alpha$ is slightly different to the value measured by Blizt \& Rosolowsky \citep{Blitz_2006}, $\alpha=0.92$. The reason for this choice is purely phenomenological -- using $\alpha=0.5$, we obtain much better agreement with the abundance of absorbers with large column densities than with $\alpha=0.92$.

We use the above procedure as our fiducial method to model the spatial distribution of neutral hydrogen in the various simulated cosmologies, and to investigate the effects induced by massive neutrinos on the spatial distribution of neutral hydrogen (see Sec. \ref{sec:neutrinos_effects}). 
To study the robustness of our SKA neutrino masses forecasts, we also model the distribution of HI using three other methods, that we describe in the following subsections.

\subsection{Pseudo-RT 2}

The HI self-shielding correction can be implemented using a method proposed by \cite{Dave_2013}. The authors of that paper state that good agreement between this method and the full radiative transfer simulations by \cite{Faucher-Giguere_2010} is achieved. It is also able to reproduce several observations, such as the HIMF \citep{Haynes_2011} at $z=0$. We will briefly describe the method here, but refer the reader to \cite{Dave_2013} for further details. 

For every gas particle in the simulation, the HI fraction in photo-ionization equilibrium is computed by taking into account the strength of the UV background and the physical properties of the particle, such as its density and temperature. Then, the HI column density, for every gas particle, is computed by integrating the SPH kernel from the radius of the particle up to a given radius (if it exists), $r_{\rm thres}$, where it reaches a given threshold that we set to $10^{17.3}~{\rm cm^{-2}}$. The method implements the HI self-shielded correction by assuming that $90\%$ of the hydrogen between $r=0$ and $r=r_{\rm thres}$ is fully neutral. Finally, the presence of molecular hydrogen is accounted for using the procedure described in the previous subsection.

\subsection{Halo-based 1}
\label{subsec:Bagla}

In contrast with the above two methods, where HI is assigned to each individual gas particle in the simulation according to its physical properties, the method depicted in this and in the next subsection are designed to assign HI to dark matter halos (see subsection \ref{subsec:low-z} for further details). These methods assume that a dark matter halo of mass $M$ at redshift $z$ hosts an amount of HI given by the deterministic function $M_{\rm HI}(M,z)$. 

\cite{Bagla_2010} proposed a parametrization of the function $M_{\rm HI}(M,z)$ as
\begin{equation}  
M_{\rm HI}(M,z) = \left\{ 
  \begin{array}{l l}	
  
    f_3(z)\frac{M}{1+M/M_{\rm max}(z)} & \quad \text{if $M_{\rm min}(z)\leqslant M$}
    \\
    0 & \quad \text{otherwise,}\\
  \end{array} \right.
\label{M_HI_Bagla3}
\end{equation} 
where the values of the parameters $M_{\rm min}(z)$ and $M_{\rm max}(z)$ correspond to halos with circular velocities equal to $v_{\rm min}=30$ km/s and $v_{\rm max}=200$ km/s at redshift $z$, respectively. The value of $f_3(z)$ is tuned to reproduce the HI density parameter $\Omega_{\rm HI}(z)$ (defined as the ratio between the comoving density of neutral hydrogen at redshift $z$ to the critical density at $z=0$). For the redshifts covered by our simulations ($3\leqslant z \leqslant5.5$), we assume that the value of $\Omega_{\rm HI}(z)$ does not depend on redshift, and that its value is equal to $10^{-3}$, both for halo-based 1 and for the halo-based 2 method. This is in excellent agreement with observations.

In practice, this method works as follows: from a simulation snapshot we identify all the dark matter halos. The total HI mass residing in a particular halo of mass $M$ at redshift $z$ is computed from Eq.~\ref{M_HI_Bagla3}. Finally, the HI mass within the halo is distributed according to some HI density profile, $\rho_{\rm HI}(M,z)$. We model the last step by splitting the total HI mass in a given halo equally amongst all the gas particles belonging to it.

In \cite{Villaescusa-Navarro_2014a}, it was shown that the above method is capable of reproducing the damped Lyman alpha absorber (DLA) column density distribution function extremely well at redshifts $z\sim[2-4]$. In \cite{Padmanabhan_2015} the authors showed that it also reproduces the HI bias at $z=0$ and the product $b_{\rm HI}(z) \times \Omega_{\rm HI}(z)$ at $z\simeq0.8$ from 21cm observations.

\subsection{Halo-based 2}

While the halo-based 1 model is capable of reproducing many observables, it does fail at reproducing one: the bias of the DLAs at $z\simeq2.3$ recently measured by the BOSS collaboration: $b_{\rm DLA}=(2.17\pm0.20)\beta_F^{0.22}$ \citep{Font_2012}, where $\beta_F$ is the Ly$\alpha$ forest redshift distortion parameter, whose value is of order 1. Here we propose a simple model that is capable of reproducing this observable as well. We propose a functional form
\begin{equation}  
M_{\rm HI}(M,z) = \left\{ 
  \begin{array}{l l}	
  
    f_4(z)M & \quad \text{if $M_{\rm min}(z)\leqslant M$}
    \\
    0 & \quad \text{otherwise,}\\
  \end{array} \right.
\label{M_HI_Paco}
\end{equation} 
where $M_{\rm min}(z)$ is chosen to be the mass of dark matter halos with circular velocities equal to $v_{\rm min}=62$ km/s at redshift $z$, and $f_4(z)$ is a parameter whose value is set to reproduce the value of $\Omega_{\rm HI}(z)$. This simple model predicts a value of the HI bias at $z=2.3$ (see Eq.~\ref{eq:bias_HI})
equal to 2.15, in perfect agreement with the observational measurements.\footnote{Note that we are making the assumption that the bias of the DLAs is the bias of the HI. This assumption is reasonable since the amount of HI in Lyman limit systems and in the Ly$\alpha$ forest is of order $\sim10\%$}

\subsection{HI at $z\textless3$}
\label{subsec:low-z}

Running all of our high-resolution hydrodynamic simulations down to $z=0$ is infeasible given the computational resources we have access to. On the other hand, the redshift range $0\leqslant z \textless 3$ may be important when forecasting the constraints on the neutrino masses that will be achievable with SKA. We therefore model the spatial distribution of neutral hydrogen at $z\textless3$ using a simple analytic model.

In \cite{Villaescusa-Navarro_2014a} it was shown that the fraction of HI outside dark matter halos in the post-reionization era is negligible. One can therefore make the assumption that all HI resides in dark matter halos. Under this assumption, and following the spirit of the halo model \citep{Cooray_Sheth_2002}, we can then predict the shape and amplitude of the HI power spectrum, in real-space, if we have the following ingredients: the halo mass function, $n(M,z)$, the halo bias, $b(M,z)$, the linear matter power spectrum $P_{\rm m}^{\rm lin}(k)$, and the functions $M_{\rm HI}(M,z)$ and $\rho_{\rm HI}(M,z)$. The functions $M_{\rm HI}(M,z)$ and $\rho_{\rm HI}(r|M,z)$ represent the average HI mass and density profile in a dark matter halo of mass $M$ at redshift $z$. 

On large, linear, scales the HI power spectrum in real-space does not depend on the $\rho_{\rm HI}(r|M,z)$ function, but only on $M_{\rm HI}(M,z)$, and it is given by
\begin{equation}
P_{\rm HI}(k,z)=b_{\rm HI}^2(z)P_{\rm m}(k,z)~,
\end{equation}
where the HI bias, $b_{\rm HI}(z)$ is given by 
\begin{equation}
b_{\rm HI}(z)=\frac{\int_0^\infty n(M,z)b(M,z)M_{\rm HI}(M,z)dM}{\int_0^\infty n(M,z)M_{\rm HI}(M,z)dM}~.
\label{eq:bias_HI}
\end{equation}
The 21cm power spectrum is given by
\begin{eqnarray}
P_{\rm 21cm}(k,z)&=&\overline{\delta T_b}^2(z)b_{\rm HI}^2(z)\left(1+\frac{2}{3}\beta(z)+\frac{1}{5}\beta^2(z)\right)\\ \nonumber
&&\times ~P_{\rm m}(k,z), 
\label{eq:P21}
\end{eqnarray}
where $\beta$ is the redshift-space distortion parameter given by $\beta(z)=f(z)/b_{\rm HI}(z)$, with $f(z)$ being the growth rate at redshift $z$. The third term on the right hand side arises from the Kaiser formula \citep{Kaiser_1987}. The value of $\overline{\delta T_b}(z)$ is given by 
\begin{equation}
\overline{\delta T_b}(z) = 189\left(\frac{H_0(1+z)^2}{H(z)}\right)\Omega_{\rm HI}(z)h~{\rm mK}~,
\label{eq:delta_Tb}
\end{equation}
where $H(z)$ and $H_0$ are the value of the Hubble parameter at redshifts $z$ and $0$, respectively. $h$ represents the value of $H_0$ in units of $100~{\rm km~s^{-1} Mpc^{-1}}$.
Also, $\Omega_{\rm HI}(z)$ does not depend on $\rho_{\rm HI}(r|M,z)$, but only on the function $M_{\rm HI}(M,z)$:
\begin{equation}
\Omega_{\rm HI}(z)=\frac{1}{\rho_{\rm c, 0}}\int_0^\infty n(M,z)M_{\rm HI}(M,z)dM~.
\label{eq:Omega_HI}
\end{equation}
The above equations make clear the central role played by the function $M_{\rm HI}(M,z)$ in studies related to 21cm intensity mapping in the post-reionization era. Modeling this function is therefore all we need to predict the shape and amplitude of the HI/21cm power spectrum on linear scales.

We use the above equations to model the 21cm power spectrum at redshifts $z\textless3$, with $M_{\rm HI}(M,z)$ given by the halo-based 1 prescription \citep{Bagla_2010}. As discussed in Sec. \ref{subsec:Bagla}, the halo-based 1 model has three free parameters: $M_{\rm min}(z)$, $M_{\rm max}(z)$ and $f_3(z)$. While the values of the parameters $M_{\rm min}(z)$ and $M_{\rm max}(z)$ are chosen to correspond to halos with circular velocities of 30 km/s and 200 km/s, the value of $f_3(z)$ is fixed by requiring that $\Omega_{\rm HI}(z)$ reproduces the observational measurements. Since at $z\textless3$ observations disfavor models with constant $\Omega_{\rm HI}$, we follow \cite{Crighton_2015} and assume a redshift dependence $\Omega_{\rm HI}(z)=4\times10^{-4}(1+z)^{0.6}$, which fully determines the value of $f_3(z)$.

For the halo mass function and halo bias we use the \cite{ST} and Sheth, Mo \& Tormen \citep{SMT} models, respectively. We use the matter power spectrum from halofit \citep{Halofit_2012} to evaluate $P_{\rm m}(k,z)$ for a given cosmological model, and compute the HI/21cm power spectra at redshifts $z=\{0, 0.25, 0.5, 0.75, 1, 1.25, 1.5, 1.75, 2, 2.5\}$.  

In cosmologies with massive neutrinos we use the CDM+baryon field, instead of the total matter density field, to compute the 21cm power spectrum (last term on the r.h.s of Eq.~\ref{eq:P21}), and to evaluate the halo mass function and halo bias \citep{Ichiki-Takada,Castorina_2014}.

We now briefly discuss the differences between applying the halo-based 1 model to our hydrodynamic simulations and the formalism we have described in this subsection. By putting the HI in the gas particles belonging to the dark matter halos of our simulations, we not only model the function $M_{\rm HI}(M,z)$, but also the HI density profile within halos, $\rho_{\rm HI}(r|M,z)$, which is ignored in the method above. The above formalism also implicitly assumes a scale-independent bias, and that the redshift-space distortions are accounted for by the Kaiser formula. The fully non-linear clustering and redshift-space distortions are taken into account by placing the HI in the simulations, however. On large, linear scales, both methods should give the same results, while on small scales the method described in this section will break down.

As such, we conclude by emphasizing that the above methodology is limited to linear scales, i.e. scales in which the HI bias is constant, and redshift-space distortions can be accounted for by using the Kaiser formula.

\section{Effect of massive neutrinos}
\label{sec:neutrinos_effects} 

Here we study the effects induced by massive neutrinos on the spatial distribution of neutral hydrogen.  We investigate how the HI abundance and clustering properties are affected by the presence of massive neutrinos by comparing the distribution of neutral hydrogen from the simulations $\nu_{\rm m}^+$ ($M_\nu\!=\!0.3$ eV) and $\nu_{\rm m}^{++}$ ($M_\nu\!=\!0.6$ eV) to the one from the fiducial simulation, $\mathcal{F}$ ($M_\nu\!=\!0.0$ eV). 

\begin{figure*}
\begin{center}
\includegraphics[width=0.9\textwidth]{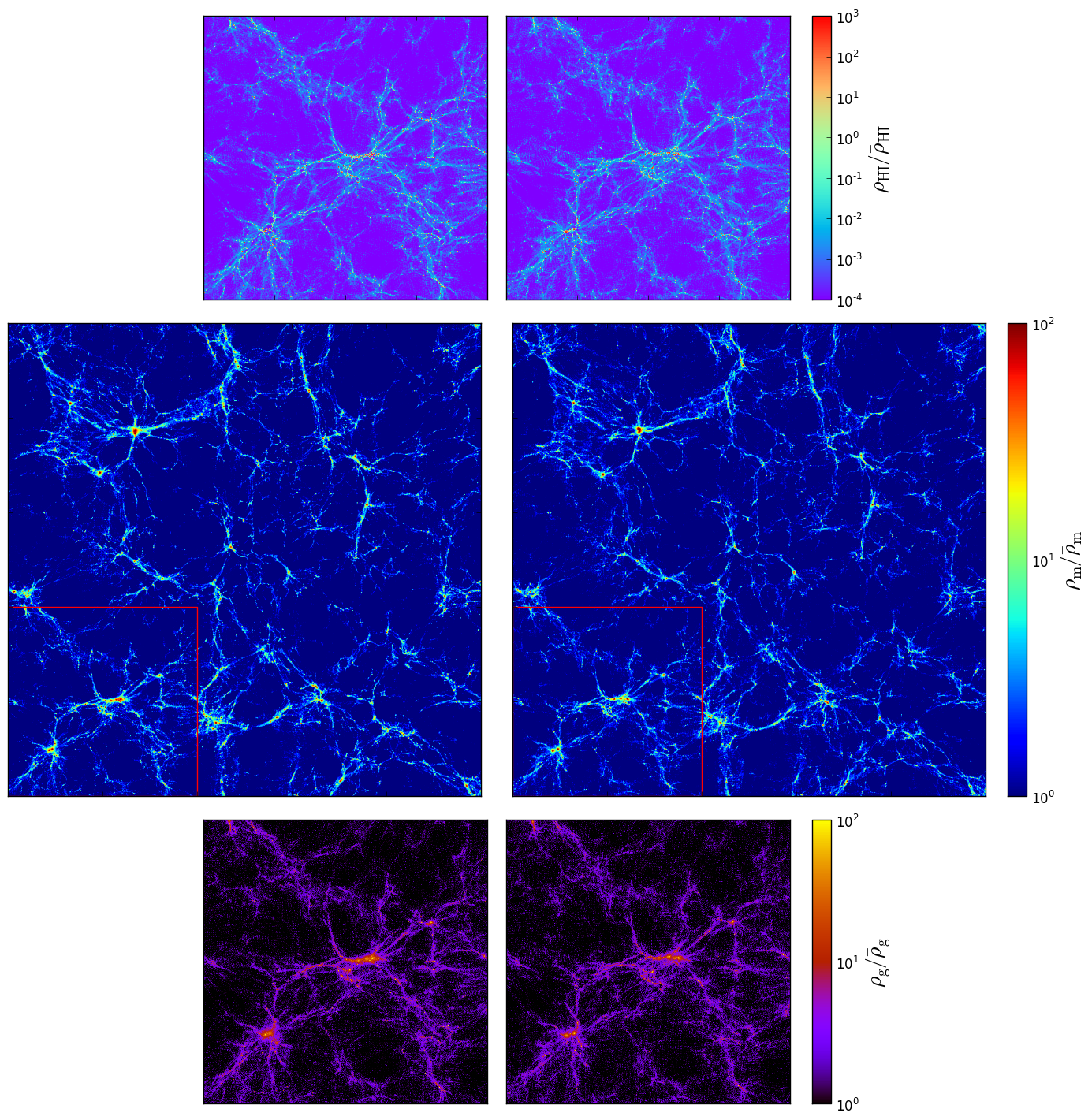}
\caption{Impact of massive neutrinos on the spatial distribution of neutral hydrogen (upper row), total matter (middle row) and gas (bottom row) at $z=3$. Panels on the left show the results for a massless neutrino cosmology while panels on the right are for a cosmological model with $M_\nu=0.6$ eV neutrinos. The middle panels display the spatial distribution of matter on in a slice of $50\times50\times2~(h^{-1}{\rm Mpc})^3$, while top and bottom panels show a zoom into the region marked with a red square (the width of those slices is also $2~h^{-1}$Mpc).}
\label{fig:Image}
\end{center}
\end{figure*}

In Fig. \ref{fig:Image} we show the spatial distribution of neutral hydrogen (top row), matter (middle row) and gas (bottom row) in a cosmology with massless neutrinos (left, simulation $\mathcal{F}$) and  with $M_\nu=0.6$ eV neutrinos (right, simulation $\nu_{\rm m}^{++}$). The images have been created by taking a slice of $2~h^{-1}$Mpc width. The spatial distribution of (total) matter is shown over the whole box (i.e. in a slice of $50\times50\times2~(h^{-1}{\rm Mpc})^3$), while the gas and HI images display a zoom over the region marked with a red square. As can be seen, the differences in the spatial distribution of matter, gas, and (in particular) neutral hydrogen between the two models are very small.

\subsection{HI abundance} 
\label{subsec:HI abundance}

\begin{figure*}
\begin{center}
\includegraphics[width=0.95\textwidth]{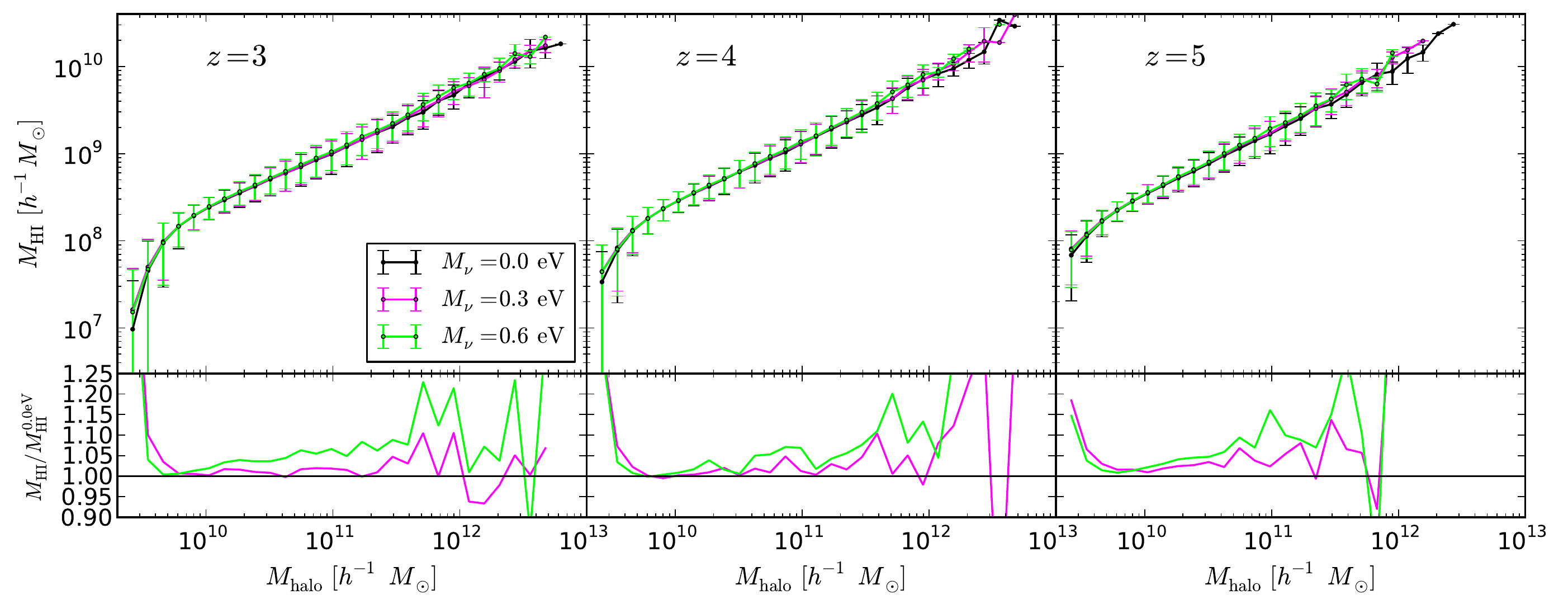}\\
\includegraphics[width=0.95\textwidth]{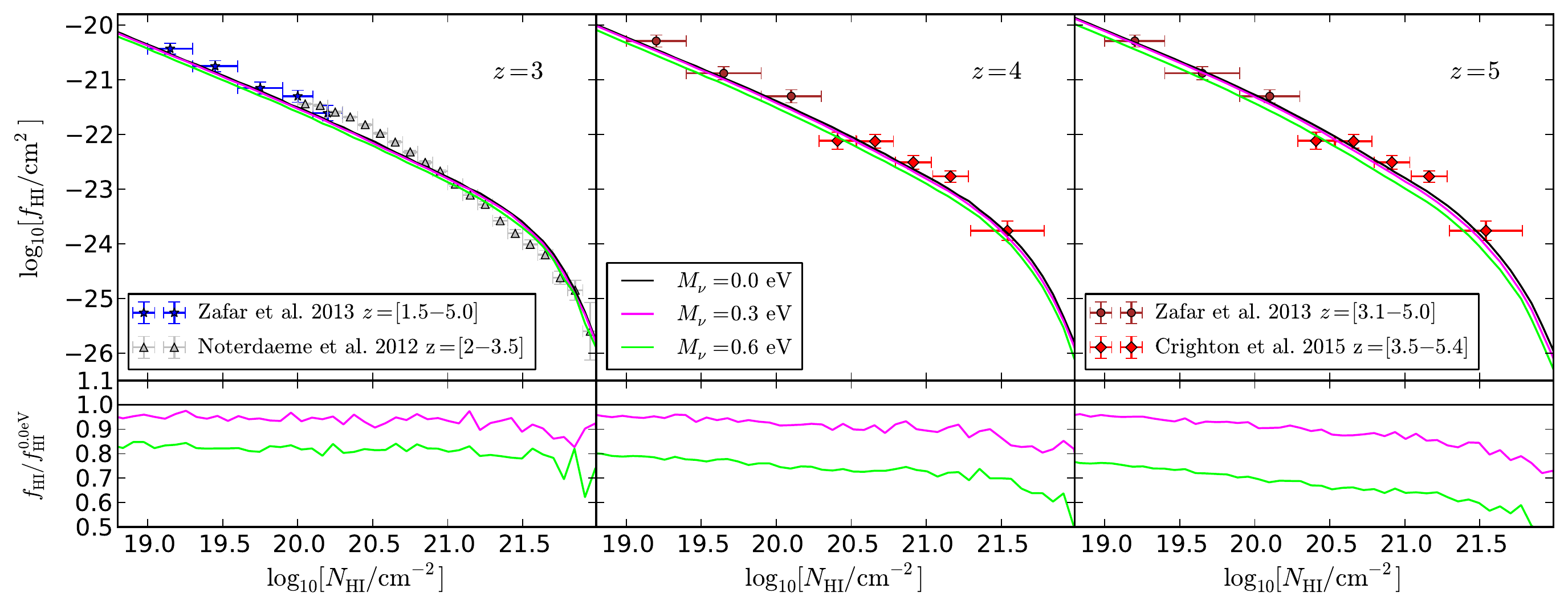}\\
\caption{\textit{Upper row:} Function $M_{\rm HI}(M)$ at redshifts $z=3$ (left), $z=4$ (middle) and $z=5$ (right) for the cosmological models with massless neutrinos (black), $M_\nu=0.3$ eV (magenta) and $M_\nu=0.6$ eV (green) when the HI is modeled using the pseudo-RT 1 method. For each dark matter halo we have computed the HI mass within it; the lines represent the running median, and error bars show the scatter around the mean. The bottom panels display the results normalized by the $M_{\rm HI}(M)$ function of the massless neutrino model. \textit{Bottom row:} Same as above but for the HI column density distribution function. The observational measurements are from \cite{Noterdaeme_2012} ($z=[2-3.5]$) and \cite{Zafar_2013} (using the whole redshift range: $z=[1.5-5.0]$) for the results at $z=3$, and from \cite{Crighton_2015} ($z=[3.5-5.4]$) and \cite{Zafar_2013} (using only the redshift range $z=[3.1-5.0]$) for the plots at $z=4$ and $z=5$.}
\label{fig:HI_abundance}
\end{center}
\end{figure*}

We now investigate how the presence of massive neutrinos impacts the function $M_{\rm HI}(M,z)$. By modeling the HI distribution using the pseudo-RT 1 method, we computed the HI mass within each dark matter halo in the simulations with massless and massive neutrinos. In the upper row of Fig. \ref{fig:HI_abundance} we show the results at redshifts $z=3,4,5$.

For a fixed dark matter halo mass, we find that halos in the massless and massive neutrino models contain the same HI mass to well within one standard deviation, although halos in the massive neutrino cosmologies do tend to host a slightly higher HI mass. The HI mass excess with respect to the massless neutrino case is $\sim7\%$ for the cosmology with 0.6 eV neutrinos, decreasing to $\sim3\%$ for the 0.3 eV cosmology, with a very weak dependence on redshift. Our results show that the HI mass excess is not uniform in mass: the most massive halos host a higher fraction of HI compared with the low mass halos.

The function $M_{\rm HI}(M,z)$ presents a cut-off at low masses, which can be seen clearly at $z=3$. This cut-off is not physical, but is due to the resolution of our simulations. In order to explore the physical cut-off arising from the fact that a minimum gas density and length is required to have self-shielded HI, simulations with higher resolution are needed. We leave this for a future work.

For the mass range accessible in our simulations, we find that the $M_{\rm HI}(M,z)$ function can be fitted by a function of the form: $M_{\rm HI}(M,z)=\left[M/M_0(z)\right]^{\alpha(z)}$. In Table \ref{tbl:M_HI} we show the best fit values of $M_0$ and $\alpha$ for the three different cosmologies at redshifts $z=3,4,5$. The value of the slope, $\alpha$, increases with redshift for all of the models, while at fixed redshift, $\alpha$ increases with the sum of the neutrino masses. This reflects a well known property: at a given redshift, the spatial distribution of matter on small scales in a cosmology with massive neutrinos is effectively younger (has grown less) than its massless neutrino counterpart \citep{Marulli_2011,Villaescusa-Navarro_2013,Castorina_2014,Costanzi_2014, Villaescusa-Navarro_2014,Massara_2014, Massara_2015}.

\begin{table}[t]
\centering{
{\renewcommand{\arraystretch}{1.6}

 \begin{tabular}{|c|c|cc|}
 \hline
$M_\nu$ & ~~$z$~~ & $M_0$ & $\alpha$ \\
(eV) & & ($h^{-1}M_\odot$) & \\
 \hline
\multirow{3}{*}{0.0} & 3 & $0.012\pm0.003$ & $0.699\pm0.006$ \\
                               & 4 & $0.015\pm0.003$ & $0.712\pm0.005$\\
                               & 5 & $0.032\pm0.008$ & $0.740\pm0.006$\\
\hline
\multirow{3}{*}{0.3} & 3 & $0.015\pm0.005$ & $0.706\pm0.007$ \\
                               & 4 & $0.031\pm0.008$ & $0.732\pm0.006$\\
                               & 5 & $0.033\pm0.006$ & $0.743\pm0.005$\\
\hline
\multirow{3}{*}{0.6} & 3 & $0.016\pm0.005$ & $0.708\pm0.007$ \\
                               & 4 & $0.027\pm0.006$ & $0.728\pm0.006$\\
                               & 5 & $0.061\pm0.016$ & $0.760\pm0.007$\\
 \hline
 \end{tabular} }}
 \caption{Best-fit parameters for $M_{\rm HI}(M,z)=(M/M_0)^\alpha$ as a function of redshift and cosmology.}\vspace{-1.5em}
 \label{tbl:M_HI}
\end{table}

\begin{figure}[t]
\includegraphics[width=\columnwidth]{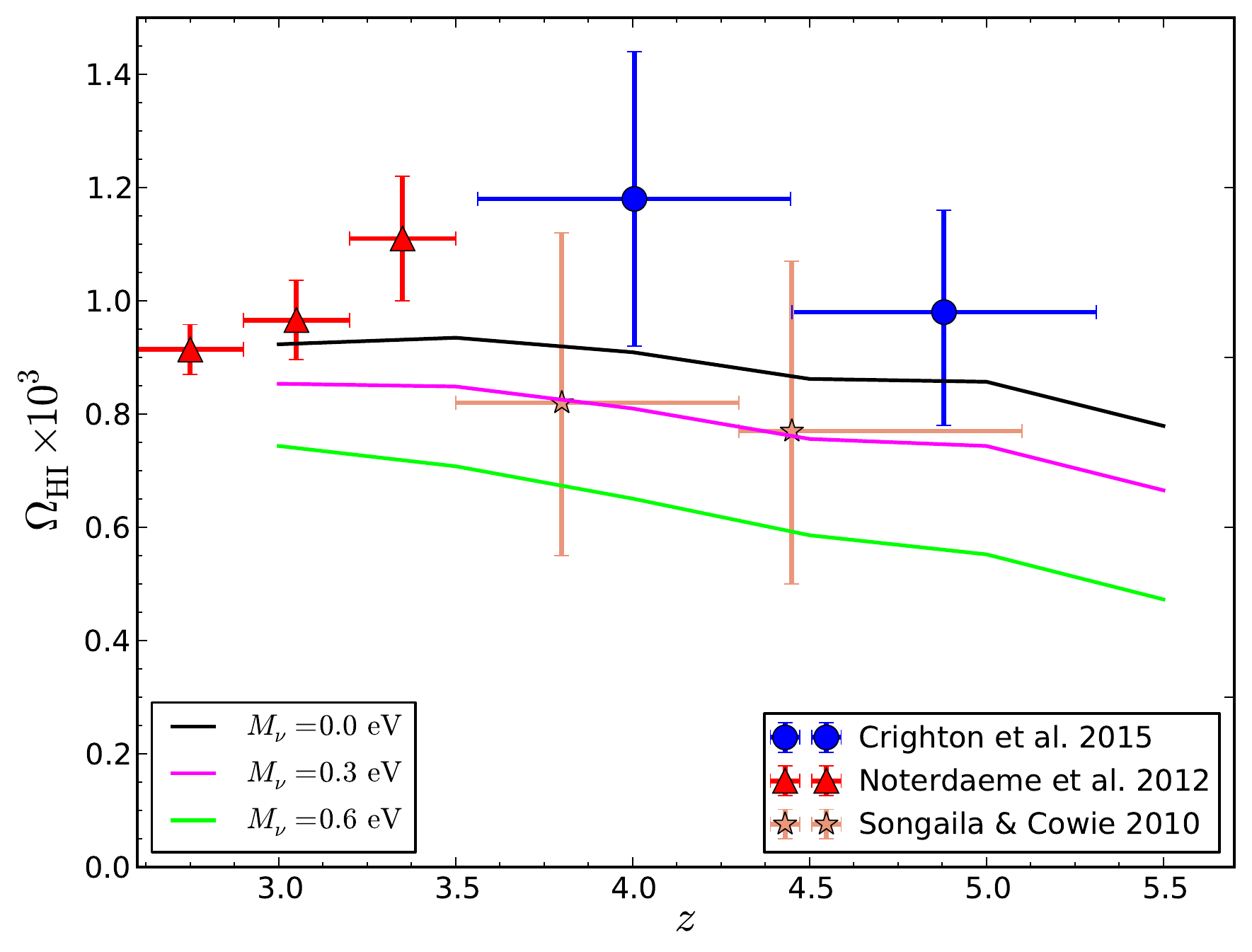}
\caption{Redshift evolution of  $\Omega_{\rm HI}$ for cosmologies with massless neutrinos (black), $M_\nu=0.3$ eV (magenta) and $M_\nu=0.6$ eV (green) when the HI is modeled using the pseudo-RT 1 method. Values obtained from the observations of \cite{Noterdaeme_2012, Crighton_2015, Songaila_2010} are shown with red, blue, and orange markers respectively.}
\label{fig:Omega_HI}
\end{figure}

Fig.~\ref{fig:Omega_HI} shows $\Omega_{\rm HI}(z)$ in each cosmology. We find that our fiducial model is capable of reproducing the values of $\Omega_{\rm HI}(z)$ obtained from the observations of \cite{Songaila_2010, Noterdaeme_2012, Crighton_2015} extremely well in the redshift range covered by our simulations. The redshift dependence of the function $\Omega_{\rm HI}(z)$ is very weak, although it is slightly more pronounced with increasing neutrino mass.

For a fixed redshift, the value of $\Omega_{\rm HI}(z)$ decreases as the sum of the neutrino masses increases. This result can be understood if we look at Eq.~\ref{eq:Omega_HI} and take into account the fact that $M_{\rm HI}(M,z)$ barely changes among cosmologies with massive and massless neutrinos. The reason for the decrement in $\Omega_{\rm HI}(z)$ is therefore due to the suppression in the abundance of halos that the presence of massive neutrinos induces \citep{Brandbyge_2010, Marulli_2011, Villaescusa-Navarro_2013, Castorina_2014, Ichiki-Takada, LoVerde_2014, Roncarelli_2015, Castorina_2015}. Using our fitting function to $M_{\rm HI}(M,z)$ from Table \ref{tbl:M_HI} (with a cut-off at $M=2\times10^{9}~h^{-1}M_\odot$), we have checked that using the massive/massless neutrino halo mass functions reproduces the decrement in $\Omega_{\rm HI}(M,z)$ induced by neutrinos.

We have also computed the HI column density distribution\footnote{See Appendix B of \cite{Villaescusa-Navarro_2014a} for a description of the procedure used to calculate this.} for the cosmologies with massless/massive neutrinos; the results are shown on the bottom row of Fig.~\ref{fig:HI_abundance}. At $z=3$ we compare our results with the measurements by \cite{Noterdaeme_2012}. While these have a mean redshift $\langle z \rangle=2.5$, they cover a redshift range $z=[2-3.5]$, and we assume no redshift evolution down to $z=3$. The abundance of DLAs in our simulations at $z=4$ and $z=5$ are compared against the recent measurements of \cite{Crighton_2015}, which have data in the redshift range $z=[3.5-5.4]$. The column density distribution function of sub-DLAs in our simulations is compared at $z=3$ against the measurements by \cite{Zafar_2013}, obtained by using the whole redshift range $z=[1.5-5.0]$, while at $z=4$ and $z=5$ we only use data in the redshift range $z=[3.1-5.0]$. We find that our fiducial model reproduces the observed abundance of DLAs and sub-DLAs very well at all redshifts.

In the bottom panels we display the ratio between the HI column density distribution function of the models with massive and massless neutrinos. We find that massive neutrinos suppress the abundance of DLAs and sub-DLAs at all redshifts, although the effect is stronger at higher redshift. This is due to the lower value of $\Omega_{\rm HI}(z)$ that is present in cosmologies with massive neutrinos. 

We conclude that, for a given mass, dark matter halos in cosmologies with massless and massive neutrinos host, on average, the same amount of neutral hydrogen. The suppression on the halo mass function induced by massive neutrinos decreases the total amount of neutral hydrogen in the Universe, given the fact that only halos above a certain mass will host HI. This manifests itself in a lower value of $\Omega_{\rm HI}(z)$, and in a deficit in the abundance of DLAs and sub-DLAs in cosmologies with massive neutrinos, with respect to the massless neutrinos model.

\subsection{HI clustering}
\label{subsec:HI_clustering}

We now investigate the impact of massive neutrinos on the clustering properties of neutral hydrogen. We focus our attention in the HI power spectrum, $P_{\rm HI}(k,z)$, the HI bias, $b_{\rm HI}(k,z)$, and the 21cm power spectrum, $P_{\rm 21cm}(k,z)$.

\begin{figure*}
\begin{center}
\includegraphics[width=0.95\textwidth]{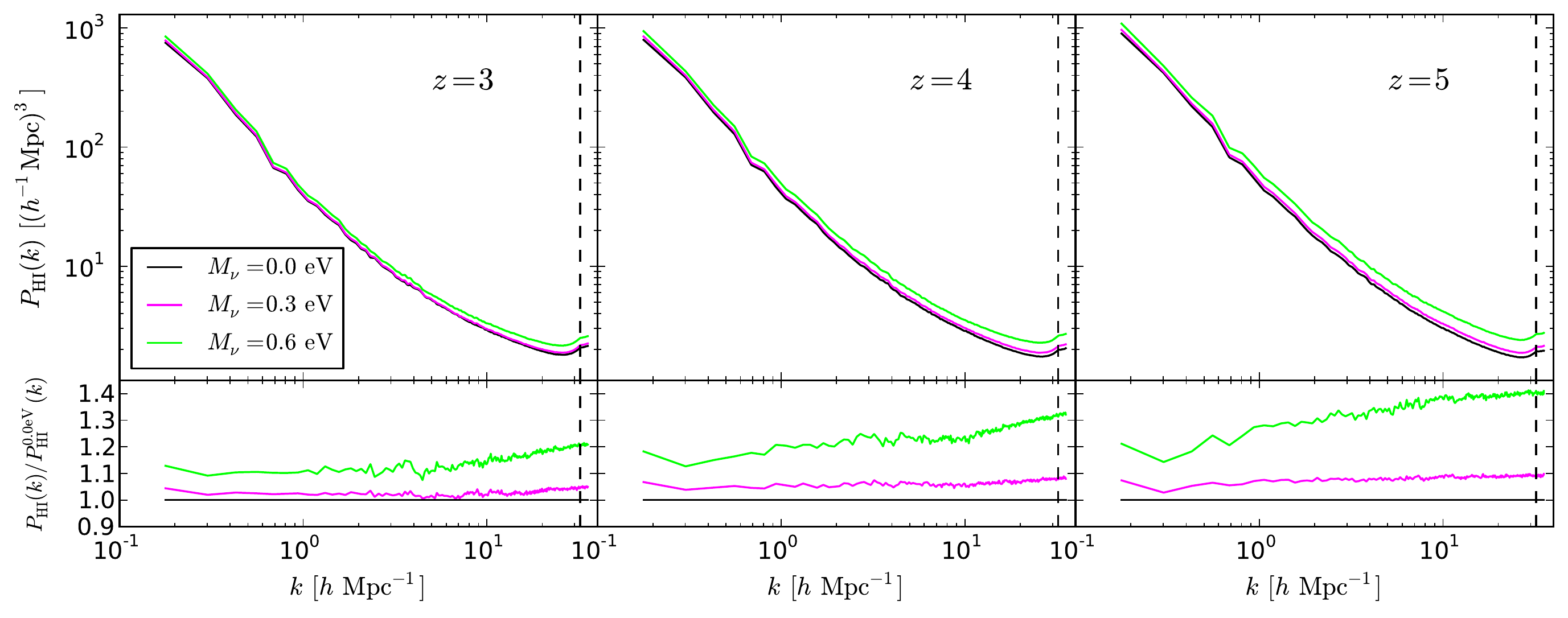}\\
\includegraphics[width=0.95\textwidth]{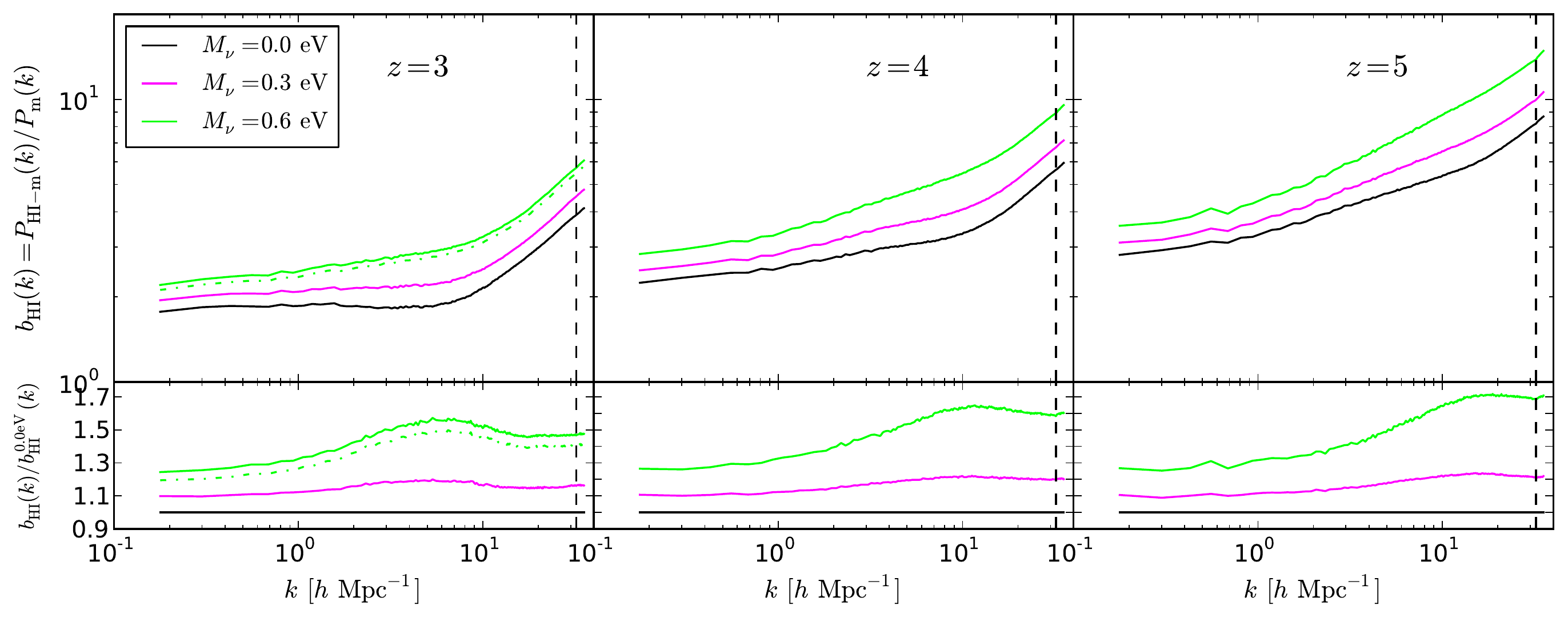}\\
\includegraphics[width=0.95\textwidth]{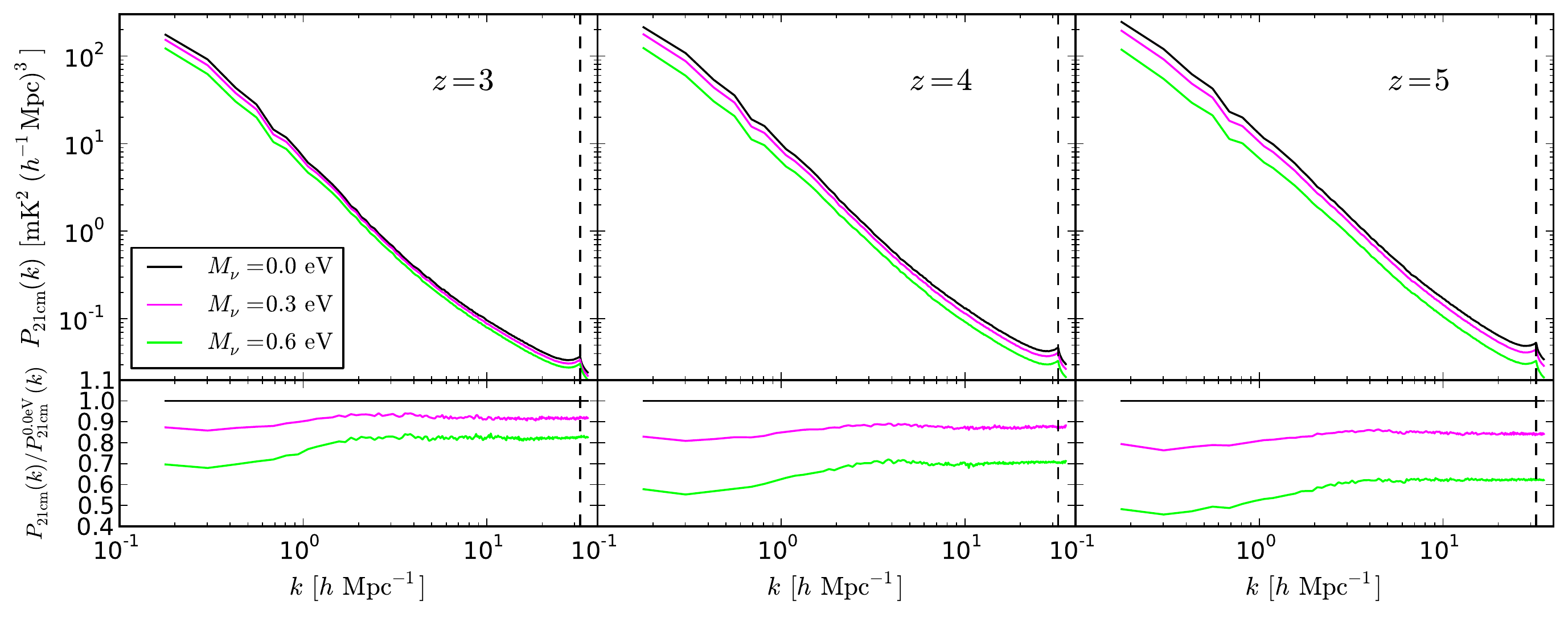}
\caption{\textit{Upper row:} Power spectrum of the neutral hydrogen field for the fiducial model with massless neutrinos (black line), the model with $M_\nu=0.3$ eV (magenta line) and the model with $M_\nu=0.6$ eV (green) at $z=3$ (left), $z=4$ (middle), and $z=5$ (right) when the HI is modeled using the pseudo-RT 1 method. The bottom panels display the ratio of the different HI power spectra to the fiducial model. The dashed vertical line represents the Nyquist frequency for our power spectrum measurements. \textit{Middle row:} same as above but for the HI bias, computed as $b_{\rm HI}(k)=P_{\rm HI-m}(k)/P_{\rm m}(k)$, with $P_{\rm HI-m}(k)$ being the HI-matter cross-power spectrum. The green dashed line is HI bias computed with respect to the CDM+baryon field (see text for details). \textit{Bottom row:} Same as above but for the 21cm field.}
\label{fig:HI_clustering}
\end{center}
\end{figure*}

In the upper row of Fig.~\ref{fig:HI_clustering} we show the HI power spectrum for the models with $M_\nu=0.0, 0.3, 0.6$ eV neutrinos at redshifts $z=3,4,5$ when the HI is modeled using the pseudo-RT 1 method. Our results show that the HI is more strongly clustered in cosmologies with massive neutrinos, with the clustering of the neutral hydrogen increasing with the sum of the neutrino masses.\footnote{Note that neutral hydrogen is also more clustered in cosmologies with warm dark matter than in the corresponding models with the standard cold dark matter, as shown in \cite{Carucci_2015}.} The bottom panels of that row show the ratios of the HI power spectra for the massive and massless neutrino cosmologies. We find that the increase in power in the massive neutrino cosmologies, relative to the fiducial model, is almost independent of scale. We find that differences in the HI clustering between models increase with redshift, however. At higher redshift, and for the $M_\nu=0.6$ eV model, the increase in power is also more scale-dependent than at lower redshift. We emphasize that the HI power spectrum, defined as $P_{\rm HI}(k,z)=\langle \delta_{\rm HI}(\vec{k},z) \delta^*_{\rm HI}(\vec{k},z) \rangle$ does not depend on the value of $\Omega_{\rm HI}(z)$, which, as we have seen above, is different for each cosmology.

We can easily see why the HI is more strongly clustered in cosmologies with massive neutrinos if we take into account that, for a fixed dark matter halo mass, the halo bias increases with the sum of the neutrino masses \citep{Marulli_2011,Villaescusa-Navarro_2014,Castorina_2014}. This happens because, as we saw above, massive neutrinos induce a suppression of the halo mass function. Thus, for a given mass, halos in cosmologies with massive neutrinos are rarer, and therefore they are more biased (for a detailed description see \cite{Massara_2014}). Since $M_{\rm HI}(M,z)$ barely changes amongst the cosmologies (or, to be more precise, increases only slightly with the sum of the neutrino masses), we should therefore expect the neutral hydrogen to be more clustered in cosmologies with massive neutrinos, as we find.\footnote{We are also implicitly assuming that the low-mass cut-off in the $M_{\rm HI}(M,z)$ function is the same mass in cosmologies with massless/massive neutrinos, which we believe is a good assumption given the fact that that function barely changes amongst the cosmologies in the mass range covered by our simulations.}

In the middle row of Fig.~\ref{fig:HI_clustering} we plot the HI bias, defined as $b_{\rm HI}(k,z)=P_{\rm HI-m}(k,z)/P_{\rm m}(k,z)$, where $P_{\rm HI-m}(k,z)$ is the HI-matter cross-power spectrum. As expected, the stronger clustering of HI in massive neutrino cosmologies is reflected in a higher HI bias in these cosmologies with respect to the fiducial, massless neutrinos model. On large scales, the HI bias increases by $\sim\!10\%$ for the model with $M_\nu=0.3$ eV neutrinos, and $\sim\!30\%$ for $M_\nu=0.6$ eV, with respect to the $M_\nu=0.0$ eV cosmology. There is a very weak dependence on redshift. On large scales, the HI bias is more scale-dependent at higher redshift, as already noted by \cite{Carucci_2015}. 

It is also worth pointing out that at $z=3$, while the HI bias is flat for wavenumbers $k\lesssim 1~h~{\rm Mpc}^{-1}$ in the $M_\nu = 0$ eV model, the models with massive neutrinos exhibit a HI bias that does depend mildly on scale. 
There are two explanations for this: 1) the bias is higher in cosmologies with massive neutrinos, and therefore more scale dependent; and 2) the HI bias, defined as above, is scale-dependent in massive neutrino cosmologies. The second option arises because it has recently been found that the halo bias in cosmologies with massive neutrinos is scale-dependent, even on very large scales \citep{Villaescusa-Navarro_2014,Castorina_2014}. In \cite{Castorina_2014} it was pointed out that if the halo bias is defined, in cosmologies with massive neutrinos, as $P_{\rm h-cb}(k)/P_{\rm cb}(k)$, where $_{\rm cb}$ denotes the cold dark matter plus baryons field, then it becomes scale-independent on large scales and universal. We have checked whether defining the HI bias as $b_{\rm HI}(k,z)=P_{\rm HI-cb}(k,z)/P_{\rm cb}(k,z)$ helps to decrease the scale-dependence of the bias at $z=3$ for the cosmologies with massive neutrinos. The middle-left panel of Fig. \ref{fig:HI_clustering} shows the results of defining the HI bias with respect to the CDM+baryons field (dashed green line). We find that by using this definition, the amplitude of the HI bias decreases because the CDM+baryon field is more strongly clustered that the total matter field in cosmologies with massive neutrinos. We do not see a significant suppression of the scale-dependence, however. We therefore conclude that HI bias is more scale-dependent in cosmologies with massive neutrinos because its value is higher \citep{Scoccimarro_2001, Cooray_Sheth_2002, Sefusatti_2005, Marin_2010}.

In the bottom row of Fig.~\ref{fig:HI_clustering} we show the 21cm power spectrum for the models with massless and massive neutrinos. The amplitude of the 21cm power spectrum is lower in massive neutrino cosmologies than in the massless neutrinos model. While this result may seem surprising, since we have seen above that HI clustering {\it increases} with the neutrino masses, it is straightforward to understand if we take into account that
\begin{equation}
P_{\rm 21cm}(k)=\overline{\delta T_b}^2(z)P_{\rm HI}^s(k),
\end{equation}
where $P_{\rm HI}^s(k)$ denotes the HI power spectrum in redshift-space, and that $\overline{\delta T_b}(z)$ depends linearly on the value of $\Omega_{\rm HI}(z)$ (see Eq.~\ref{eq:delta_Tb}). In other words, the amplitude of the 21cm power spectrum not only depends on the HI clustering, but also on the value of $\Omega_{\rm HI}(z)$. Therefore, even if the HI is more clustered in cosmologies with massive neutrinos, the lower value of $\Omega_{\rm HI}(z)$ in those cosmologies drives the suppression of power we find in the 21cm power spectrum.

In contrast with what happens with the HI power spectrum, the ratio of the 21cm power spectra in the massive and massless neutrinos cosmologies exhibits a characteristic dependence on scale. On very small scales the ratio is flat, while it decreases around $k\sim\!2\!-\!3~h\,{\rm Mpc}^{-1}$. The ratio appears to reach a minimum around $k\!\sim\!0.3~h\,{\rm Mpc}^{-1}$, although more realizations would be needed to confirm that this is not an artefact of sample variance.

We conclude that the HI is more strongly clustered in cosmologies with massive neutrinos. This is because, for a fixed halo mass, the halo bias and HI mass increase with the sum of the neutrino masses. On the other hand, the amplitude of the 21cm power spectrum decreases as the neutrino masses increase because it depends on the total amount of neutral hydrogen, $\Omega_{\rm HI}(z)$, which is lower in massive neutrino cosmologies.

\section{SKA forecasts} 
\label{sec:forecasts} 

In this section we present Fisher forecasts for the neutrino mass constraints that will be achievable with measurements of the 21cm power spectrum from future intensity mapping experiments. We consider two surveys, both with Phase 1 of the Square Kilometre Array (SKA): a wide and deep survey at low redshift ($0 \lesssim z \lesssim 3$), using the SKA1-MID array, and a narrow and deep survey at higher redshift ($3 \lesssim z \lesssim 6$) using SKA1-LOW. The forecasts for the latter use 21cm power spectra measured from our high-resolution hydrodynamic simulations, which extend into the non-linear regime, while the former use only linear modes from power spectra derived using halofit \citep{Halofit_2012}.

In all cases we include a number of nuisance parameters related to the method used to model the spatial distribution of neutral hydrogen, and introduce conservative priors from contemporary experiments to break degeneracies between cosmological parameters. We also consider more pessimistic choices of instrumental parameters than given in the baseline designs of the SKA1 sub-arrays. The result is a conservative set of forecasts for the neutrino mass constraints that one can expect to obtain with the SKA. Nevertheless, a number of systematic effects could further degrade the constraints; we assess their likely importance at the end of this section.

\subsection{Fisher matrix formalism}

To produce our forecasts, we adopt the Fisher matrix formalism for 21cm surveys that was described in \cite{Bull_2015}. The Fisher matrix can be derived from a Gaussian expansion of the likelihood for a set of 21cm brightness temperature fluctuation maps about a fiducial cosmological model. For a single redshift bin, the Fisher matrix can be written as
\be \label{eq:fisher}
F_{ij} = \frac{1}{2} \int \frac{d^3k}{(2\pi)^3} V_{\rm eff}(\vec{k}) \frac{\partial \log P_{\rm 21cm}(k)}{\partial \theta_i} \frac{\partial \log P_{\rm 21cm}(k)}{\partial \theta_j},
\ee
where $\{\theta\}$ is a set of cosmological and astrophysical parameters to be constrained or marginalised over, and we have neglected cosmological evolution within the redshift bin. The effective survey volume is given by
\be
V_{\rm eff}(\vec{k}) = V_{\rm phys} \left ( \frac{P_{\rm 21cm}}{P_{\rm 21cm} + P_{N}} \right )^2,
\ee
with $V_{\rm phys}$ the comoving volume of the redshift bin. The measured power spectrum is a combination of the cosmological 21cm brightness temperature power spectrum and the instrumental noise, $P_{\rm tot} = P_{\rm 21cm} + P_{N}$, where we have assumed the noise to be Gaussian and uncorrelated, as well as uncorrelated with the signal. These are reasonable assumptions in the absence of more detailed instrumental simulations, and if shot noise is negligible \citep[this should be the case; see][]{Santos:2015bsa}.

The form of the noise power spectrum depends on the type of radio telescope used to observe the 21cm emission. The basic division is between {\it interferometers}, which coherently cross-correlate the signals from pairs of receivers to observe certain Fourier modes on the sky (with mode wavelength corresponding to the inverse separation of the receivers), and {\it autocorrelation experiments}, which construct sky maps pixel-by-pixel from the detected (autocorrelation) signals from individual receivers for many different pointings. The relative merits of the two types are discussed in \cite{Bull_2015}, where models for their noise power spectra are also derived. We will simply quote them here. The basic expression is
\be
\frac{P_{N}}{r^2 r_\nu} = \frac{T^2_{\rm sys} S_{\rm area}}{t_{\rm tot} \nu_{\rm 21cm}} 
B^{-1}_\parallel B^{-2}_\perp \mathcal{I},
\ee
where $T_{\rm sys}$ is the system temperature, $S_{\rm area}$ is the total survey area, $t_{\rm tot}$ is the total observing time, $\nu_{\rm 21cm}$ is the rest frame 21cm emission frequency ($\approx 1420$ MHz), and $r_\nu = c\,(1+z)^2 / H(z)$.
The system temperature is approximately the sum of the sky temperature, $T_{\rm sky} \approx 60\, \mathrm{K}\, (\nu / 300 {\rm MHz})^{-2.5}$, and the instrumental temperature, $T_{\rm inst}$. There is an effective beam in the radial (frequency) direction due to the frequency channel bandpass, which we model as
\be
B_\parallel = \exp{\left (-\frac{(k_\parallel r_\nu \delta\nu)^2}{16\log 2\, \nu_{\rm 21}^2}\right )}, \nonumber
\ee
where $\delta \nu$ is the channel bandwidth (assumed to be 100 kHz), and $k_\parallel$ is the Fourier wavenumber in the radial direction. The remaining factor of $B^{-2}_\perp \mathcal{I}$ describes the sensitivity as a function of angular scale, and is given by
\bea
&{\rm FOV} \big / n(d) & ~~~~{\rm (interferometer)} \nonumber\\
&\exp\left( \frac{(k_\perp r\, \theta_{\rm B})^2}{16 \log 2} \right) \frac{1}{N_d N_b} & {\rm ~~~~(autocorrelation)}. \nonumber
\eea
For interferometers, the important factors are the instantaneous field of view (${\rm FOV}$) and the baseline density distribution, $n(d)$, where $d$ is the baseline length (related to the transverse Fourier wavenumber by $k_\perp r = 2\pi d / \lambda$). For autocorrelation, we have assumed a Gaussian beam of FWHM $\theta_{\rm B}$ for each element of an array with $N_d$ dishes, and $N_b$ beams per dish.

The model for the 21cm (signal) power spectrum depends on the range of redshifts and physical scales in question. Our high-resolution simulations, described above, cover significantly non-linear scales at $z \lesssim 5.5$, but cannot be extended into the linear regime or beyond $z=3$ due to resolution requirements. For $z \ge 3$, we use the pseudo-RT 1 model as our fiducial HI model. We take the 21cm power spectrum measured from the simulations at each redshift, and then extrapolate it to larger scales using the halofit matter power spectrum, multiplied by a (scale-independent) bias that matches the amplitude of the spectra at $k = 0.2\, h$Mpc$^{-1}$. These scales are sufficiently linear for the halofit power spectrum to be a good approximation, and the bias has become almost scale-independent by this $k$ value at all relevant redshifts, as shown in Fig.~\ref{fig:HI_clustering}.

At redshifts $z < 3$, we restrict ourselves to approximately linear scales only, $k \le 0.2\, h$Mpc$^{-1}$. The power spectra are calculated by multiplying the halofit matter power spectrum by a bias function and brightness temperature model derived from the halo-based 1 prescription for the function $M_{\rm HI}(M,z)$; see Section \ref{subsec:low-z}. 

In all cases, we assume that only the monopole of the redshift-space 21cm power spectrum can be recovered. In principle, one could use the full redshift-space power spectrum, which would also allow constraints on the linear growth rate to be extracted, but we forego this to keep our analysis conservative, and because the higher order multipoles extracted from the high redshift simulations are noisier than the monopole, making it more difficult to numerically differentiate them reliably.

We calculate the derivatives of $\log P_{\rm 21cm}$ required by Eq.~(\ref{eq:fisher}) numerically, using central finite differences: $\partial f/\partial x \approx [f(x+\Delta) - f(x-\Delta)] / 2\Delta$, where $\Delta$ is the finite difference step size and all other parameters are held fixed at their fiducial values. While central differences require twice as many simulated spectra as forward differences, they are more accurate for sufficiently small step sizes. The step size chosen for each parameter can be found from the list of simulations in Table \ref{tab_sims}.

\subsection{Survey specifications}

Most current and planned 21cm intensity mapping experiments are optimised either for detecting the baryon acoustic oscillations at $z \sim 1$, or studying the epoch of reionization at $z \gtrsim 6$. While it is not a dedicated IM experiment, Phase 1 of the Square Kilometre Array (SKA) will be able to cover the entire redshift range of interest here -- from $z=0$ to $z \sim 6$ -- so we adopt it as the reference experiment for our forecasts.

SKA1 will consist of two sub-arrays: SKA1-MID, a mid-frequency dish array to be situated in South Africa, and SKA1-LOW, an array of low-frequency dipole antennas that will be constructed in Western Australia. SKA1-MID will support receivers that cover several different bands, spanning the frequency range from $350$ MHz up to $\sim14$ GHz. Only Bands 1 and 2 are relevant for detecting redshift 21cm emission; these are currently planned to cover the ranges $350 - 1050$ MHz and $900 - 1670$ MHz respectively. For simplicity, however, we will assume a single band from $350 - 1420$ MHz with the specifications of Band 1 (summarized in Table \ref{tbl:specs}).

The MID array is expected to consist of approximately $200$ dishes of diameter $15$m, primarily designed for use as an interferometer. It will be able to perform IM surveys much more efficiently if used in an autocorrelation mode, however \citep[assuming that technical issues such as the effects of correlated noise can be mitigated; see][]{Bull_2015, Santos:2015bsa}, so we will assume an autocorrelation configuration in our forecasts. With this setup, a total survey area of $\sim\! 25,000$ deg$^2$ should be achievable for a 10,000 hour survey; we take these values as our defaults in what follows, but also study the effect of changing $S_{\rm area}$ in Section \ref{sec:surveydependence}.

SKA1-LOW will cover the band $50 \le \nu \le 350$ MHz ($3 \le z \le 27$), making it one of the few proposed IM experiments that overlaps with the redshifts of our simulations, $3 \lesssim z \lesssim 6$. (The Murchison Wide Field array\footnote{\url{http://www.mwatelescope.org/}} also partially covers this range, $80 \le \nu \le 300$ MHz, $3.7 \le z \le 16.8$.) We use only the high frequency half of the band, and modify the maximum frequency slightly to $375$ MHz in our forecasts, to better fit with the redshift coverage of our simulations. We use a baseline distribution based on the description in \cite{dewdney2013ska1}. All other specifications are given in Table \ref{tbl:specs}.

The instantaneous field of view of SKA1-LOW is 2.7 deg$^2$ at $z=3$, increasing to 6 deg$^2$ at $z=5$. As our default, we consider a deep 10,000 hour survey over 20 deg$^2$. This is comparable to what will be required to detect the 21cm power spectrum from the epoch of reionization (EoR). A proposed `deep' EoR survey with SKA1-LOW will perform $1000$ hour integrations in the band $50-200$ MHz at five separate pointings on the sky, for example, with each pointing covering $\sim\! 20$ deg$^2$ at the reference frequency 100 MHz \citep{AASKA_EOR}. These pointings cover a total survey area of $\sim 20$ deg$^2$ at the centre of our chosen SKA1-LOW band ($\approx 290$ MHz), so the surveys could be performed `commensally'. We assume double the total survey time, however; while the EoR survey may be limited to observing only on winter nights to mitigate ionospheric and foreground contamination \citep{AASKA_EOR}, these effects are less of a restriction at higher frequencies. The dependence of our results on $S_{\rm area}$ and $t_{\rm tot}$ is studied in Section \ref{sec:surveydependence}.

\begin{table}[t]
\centering{
{\renewcommand{\arraystretch}{1.6}
 \begin{tabular}{|l|cc|}
 \hline
 & {\bf SKA1-LOW} & {\bf SKA1-MID} \\
 \hline
 $T_{\rm inst}$ [K] & $40 + 0.1 T_{\rm sky}$ & 28 \\
 $N_d \times N_b$ & $911 \times 3$ & $190 \times 1$ \\
 $\nu_{\rm min}$ [MHz] & 210 & 375 \\
 $\nu_{\rm max}$ [MHz] & 375 & 1420 \\
 $A_{\rm eff}(\nu_{\rm crit})$ [m$^2$] & 925 & 140 \\
 $S_{\rm area}$ [deg$^2$] & 20 & 25,000 \\
 $t_{\rm tot}$ [hrs] & 10,000 & 10,000 \\
 \hline
 \multirow{3}{*}{$z$ bin edges} & \multicolumn{1}{l}{2.75, 3.25, 3.75,} & \multicolumn{1}{l|}{0, 0.125, 0.375, 0.625,} \\
  & \multicolumn{1}{l}{4.25, 4.75, 5.25,} & \multicolumn{1}{l|}{0.875, 1.125, 1.375,} \\
  & \multicolumn{1}{l}{5.75} & \multicolumn{1}{l|}{1.625, 1.875, 2.2, 2.8} \\
 \hline
 \end{tabular} }}
 \caption{Array specifications assumed in our forecasts. More detailed specifications are given in \cite{Santos:2015bsa}.}
 \label{tbl:specs}
\end{table}

\subsection{Model uncertainties and nuisance parameters}\label{sec:nuisance}

As mentioned above, the neutrino mass constraints depend on being able to accurately measure the shape and amplitude of the 21cm power spectrum, also as a function of redshift. While the power spectra derived from our simulations are self-consistent given a particular HI model, most models are calibrated off current, imperfect data, and can be inconsistent with one another. Our forecasts are therefore subject to a number of modelling uncertainties, which we attempt to take into account in this section. (We also explore the dependence of our forecasts on the assumed fiducial model in Section \ref{sec:hidependence}.)

\begin{table*}[t]
\centering{
{\renewcommand{\arraystretch}{1.6}
\resizebox{17cm}{!}{
 \begin{tabular}{|l|c|c|c|}
 \hline
 \multirow{3}{*}{\bf Massive neutrino datasets} & \multicolumn{3}{c|}{$\sigma(M_\nu)$ / eV (95\% CL)} \\
 \cline{2-4}
  & ~~~~~~~~~~~~~~~~~~~~ & \multicolumn{1}{l|}{+ Planck CMB} & \multicolumn{1}{l|}{+ Planck CMB} \\
  & &  & \multicolumn{1}{l|}{+ Spectro-z} \\
 \hline
 Planck $M_\nu$                       & ---   & 0.461 & 0.094 \\
 \hline
 SKA1-LOW                             & 0.311 & 0.208 & 0.118 \\
 SKA1-MID                             & 0.268 & 0.190 & 0.104 \\
 SKA1-LOW + SKA1-MID                  & 0.183 & 0.145 & 0.082 \\
 \hline
 SKA1-LOW + Planck $M_\nu$            & ---   & 0.089 & 0.076 \\
 SKA1-MID + Planck $M_\nu$            & ---   & 0.071 & 0.065 \\
 SKA1-LOW + SKA1-MID + Planck $M_\nu$ & ---   & 0.067 & 0.058 \\
 \hline
 \end{tabular} }
}}
 \caption{Marginal 2$\sigma$ (95\% CL) constraints on the neutrino mass, for various combinations of surveys and prior information.}
 \label{tbl:marginal}
\end{table*}

An important but poorly-constrained contribution to the HI power spectrum is the HI bias, which is a function of both redshift and scale. To take into account the uncertainty associated with the bias model, we incorporate a simple template-based bias parametrisation into our signal model. This is constructed by first measuring the bias from the simulations as a function of scale and redshift, $b(z, k)$. We then introduce an amplitude, $A$, and shift parameter, $\alpha$, into the definition of the monopole of the redshift-space 21cm power spectrum in each redshift bin, such that
\bea
P^{0}_{\rm 21cm}(z_i, k) &=& \overline{\delta T_b}^2(z_i) \left [ A_i\, b_{\rm fid}(z_i, \alpha_i k) \right ]^2 \\
&&\times \left [ 1 + \frac{2}{3}\beta_{\rm fid} + \frac{1}{5}\beta_{\rm fid}^2 \right ] P^{\rm fid}_{\rm m}(z_i, k). \nonumber
\eea
We marginalise over the amplitudes in all of our forecasts, but the shift parameters are only marginalised for the high-redshift survey; the low-redshift forecasts use only linear scales by construction, where the bias has no scale-dependence, so the $\{\alpha_i\}$ are unconstrained. This marginalisation procedure is pessimistic in that it assumes no prior constraints on the redshift evolution of the HI bias; $A_i$ is left completely free in each bin, while there are in fact some existing constraints on its value. The $A_i$ parameter also absorbs the uncertainty in other parameters that affect the normalization, such as $\beta$, which we do not attempt to constrain separately. We have assumed somewhat more about the scale dependence -- the overall shape of the function is fixed -- but by allowing the shift parameter (and thus the scale at which scale-dependence kicks in) to be free in each redshift bin, we are still being reasonably conservative with our model.

\subsection{Prior information}\label{sec:prior}

Without additional information about the cosmological model parameters, the Fisher matrix can suffer from degeneracies, severely degrading the forecast constraints. As well as presenting results for the SKA1-LOW and MID surveys alone, we also include Fisher matrix priors for two datasets that will be available contemporary with SKA1: a full-sky CMB experiment (Planck), and a future spectroscopic galaxy redshift survey (e.g. Euclid or DESI).

For the CMB prior, we used an approximate Fisher matrix based on the 2015 Planck-only temperature + polarisation constraints on a two-parameter ($M_\nu + N_{\rm eff}$) extension to $\Lambda$CDM.\footnote{We used the \texttt{base\_nnu\_mnu\_plikHM\_TT\_lowTEB} MCMC chains; $N_{\rm eff}$ is the effective number of relativistic species.} To construct it, we took the public Planck MCMC chains for this model, calculated the covariance matrix for all parameters of interest, and then inverted it. This procedure discards the non-Gaussian part of the posterior distribution, which enforces consistency with the Gaussianity assumption of Fisher forecasting but loses some information compared with the actual Planck constraints. Considering the other caveats and simplifications inherent in Fisher forecasting, and the approximate Gaussianity of most subspaces of the posterior distribution, this is a reasonable approximation. Note that this method results in a prior that has been implicitly marginalised over a large number of systematic and instrumental effects, such as foregrounds and calibration errors. We also marginalised over the effective number of relativistic degrees of freedom, $N_{\rm eff}$, assuming that only the CMB gives information on this parameter. 

For the galaxy redshift survey prior, we performed separate Fisher forecasts for a future spectroscopic survey of $\sim 6 \times 10^7$ galaxies over 15,000 deg$^2$ from $0.65 \lesssim z \lesssim 2.05$, as was also used in \cite{Bull_2015}. This is a similar specification to the planned Euclid mission, and has comparable properties to DESI. Following \cite{Amendola:2012ys}, we took the bias to be $b(z) = \sqrt{1 + z}$, marginalised as a free parameter in each redshift bin of width $\Delta z = 0.1$. We also assumed that the information from the broadband shape of the power spectrum and redshift-space distortions can be used, and marginalised over a non-linear damping of small scales in redshift-space, $\sigma_{\rm NL}$. Conservatively, we assumed that no information on the neutrino mass is obtained from the galaxy survey. When combining the galaxy survey Fisher matrix with the low-redshift SKA1-MID IM Fisher matrix, we also assumed that the surveys are independent, although in reality they both probe the same underlying matter density field. We do not expect the effect of partially-overlapping survey volumes to be large (and note that the SKA1-LOW survey volume has no overlap).

With the priors included, we are therefore forecasting for the following parameter set:
\bea
\{ h, \Omega_c, \Omega_b, A_s, n_s, M_\nu \}_{\rm base} + \{ A_i, \alpha_i \}_{\rm bias} + \{ b_j, \sigma_{\rm NL} \}_{\rm gal.} \nonumber
\eea

\subsection{Neutrino mass constraints}

\begin{figure*}[t]
\includegraphics[width=2\columnwidth]{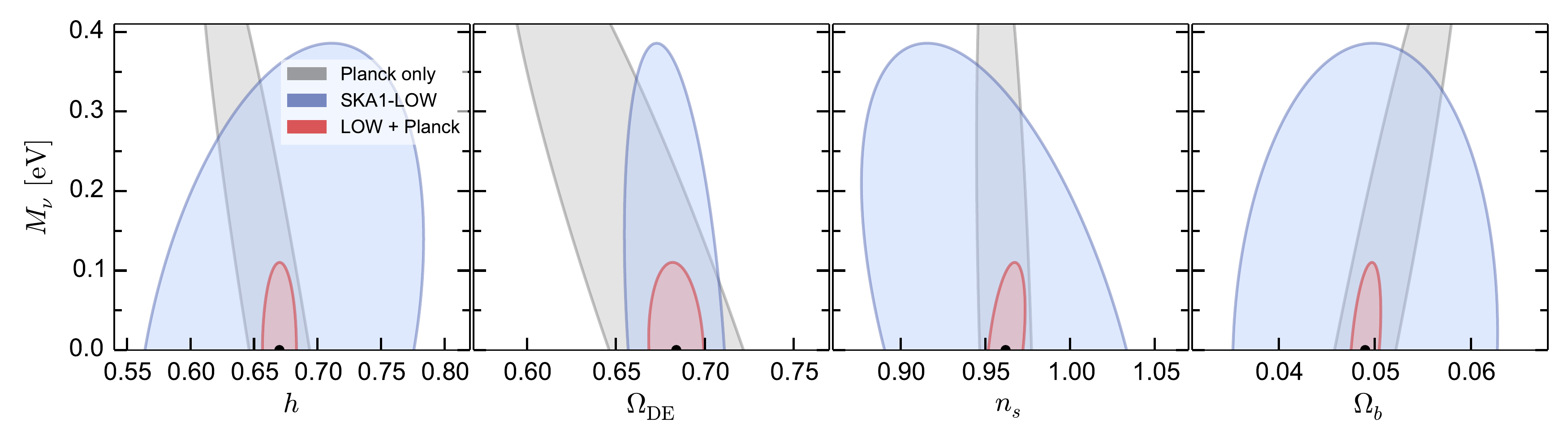}
\includegraphics[width=2\columnwidth]{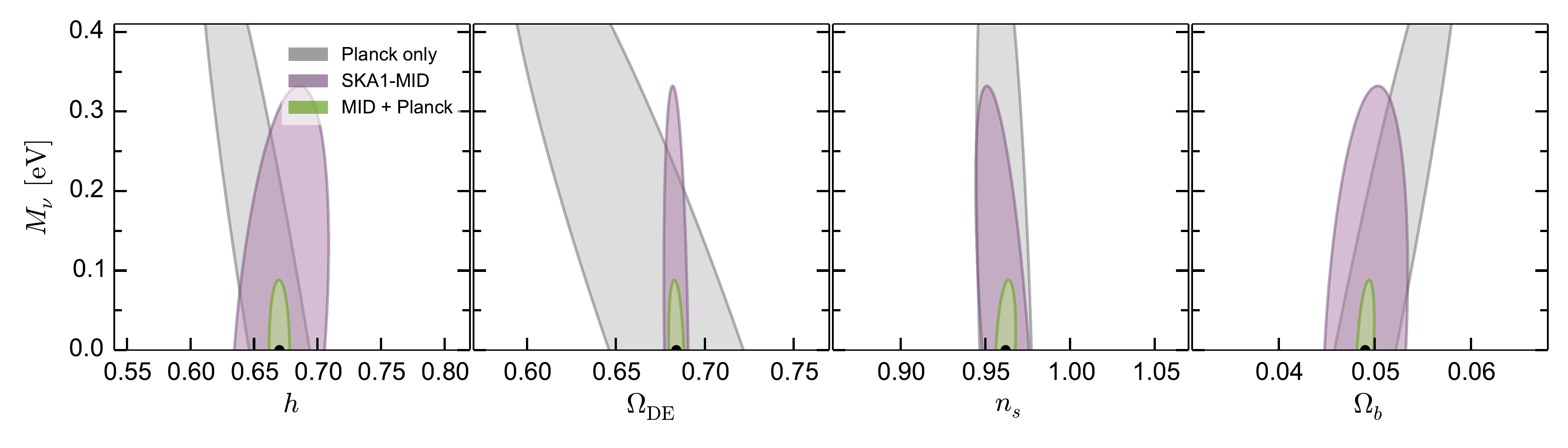}
\caption{Forecast 2$\sigma$ marginal constraints on $M_\nu$ and several cosmological parameters, for various combinations of surveys. SKA1-MID constrains the cosmological parameters significantly better than LOW, but the neutrino mass constraints are comparable.}
\label{fig:correlations}
\end{figure*}

\begin{figure}[t]
\includegraphics[width=\columnwidth]{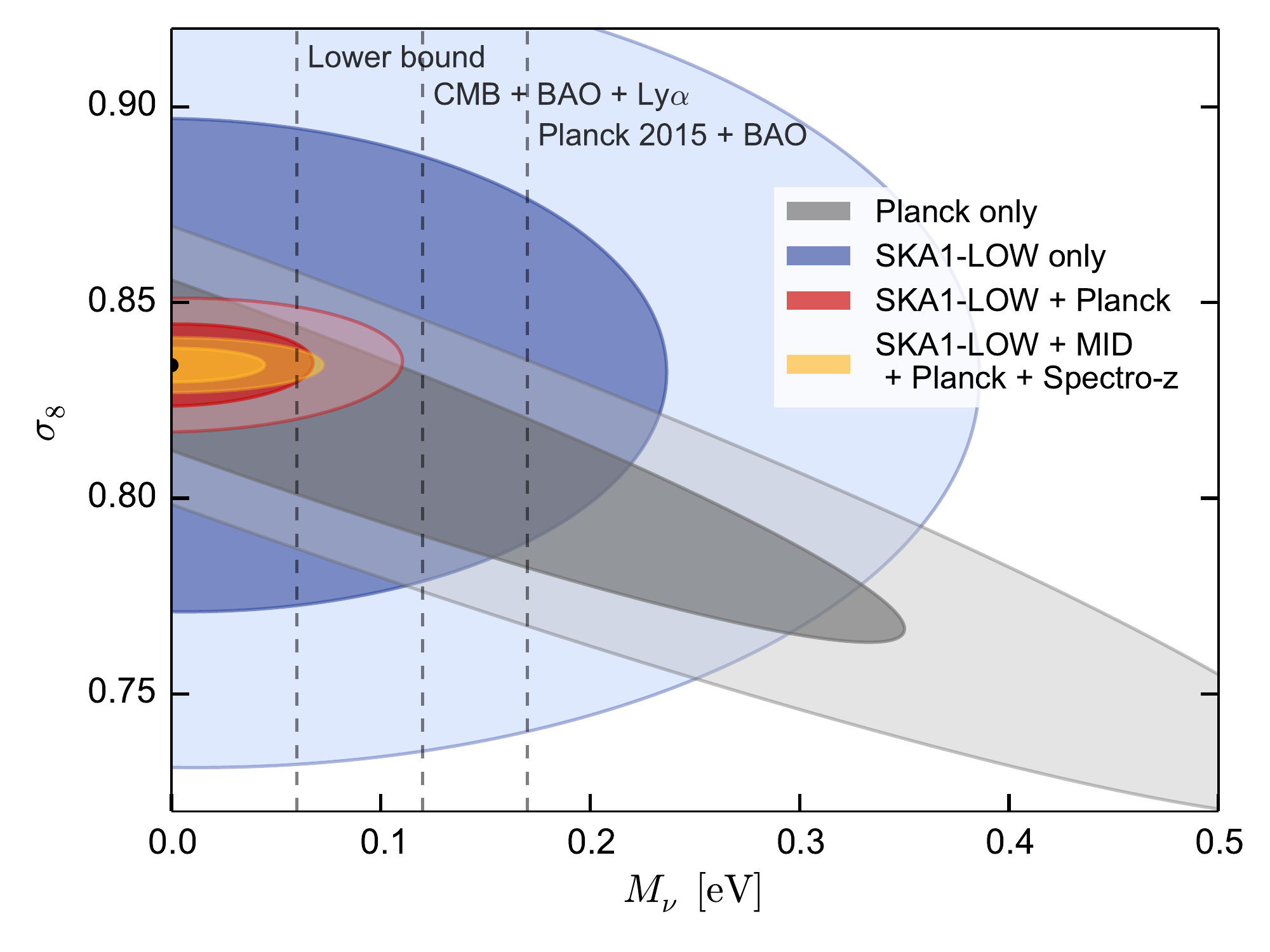}
\caption{Forecast marginal 1- and 2$\sigma$ constraints on $M_\nu$ and $\sigma_8$, for SKA1-LOW and Planck, and for the combination of all experiments considered here. The combined constraint (red) is much better than the individual constraints because multiple parameter degeneracies are broken by combining the datasets. The lower bound, CMB+BAO+Ly$\alpha$, and Planck 2015 95\% limits are shown as vertical dashed lines from left to right respectively.}
\label{fig:mnu_sig8}
\end{figure}

The results of our forecasts are summarized in Table \ref{tbl:marginal} and Figs.~\ref{fig:correlations} and \ref{fig:mnu_sig8}. All forecasted errors are quoted at the 2$\sigma$ (95\%)  CL.

For our approximate Planck 2015 Fisher matrix, the constraint on $M_\nu$ is 0.46 eV, which is consistent with the actual Planck 2015 temperature + polarisation upper limit of 0.49 eV \citep{Planck_2015}.\footnote{These figures can be compared with pre-data release forecasts for Planck \citep[e.g.][]{2003PhRvL..91x1301K, Perotto:2006rj, Kitching:2008dp}, which predicted $\sigma(M_\nu) \sim 0.3 - 0.9$ eV (95\% CL).} Our forecasts suggest that intensity mapping surveys with SKA1-MID and LOW will be able to moderately surpass the Planck constraints, yielding $\sigma(M_\nu) \approx 0.3$ eV without the addition of any prior information to break the strong correlations that exist between some of the cosmological parameters for the IM surveys. This improves to $\approx 0.2$ eV when LOW and MID are combined. Adding a Planck prior on the cosmological parameters only (i.e. ignoring information on $M_\nu$ from the CMB data) improves the constraint slightly, reaching $\sim 0.15$ eV for MID + LOW, but it is the inclusion of spectroscopic redshift survey data that most strongly breaks the cosmological parameter degeneracies; for a Planck CMB + spectro-z prior, we expect $\sigma(M_\nu) \approx 0.08$ eV for SKA1-LOW + MID.

Similar precision can be gained by combining either of the IM surveys with the full Planck data (including the neutrino mass constraint), however, without the need to add the spectroscopic galaxy survey: we obtain $\sigma(M_\nu) = 0.067$ eV for SKA1-LOW + MID + Planck. This is a result of the IM and CMB constraints having different correlation directions for the covariances between $M_\nu$ and certain cosmological parameters -- Fig.~\ref{fig:mnu_sig8} shows how the IM and CMB data have somewhat complementary correlation directions in the $M_\nu-\sigma_8$ plane, for example, and Fig.~\ref{fig:correlations} shows correlations with other parameters. This complementarity is also present in other types of large-scale structure survey \citep{Carbone:2010ik, 2011APh....35..177A}. Adding the spectroscopic survey data improves the constraints on other cosmological parameters somewhat, resulting in a slight further improvement to $\sigma(M_\nu) = 0.058$ eV for the combination of SKA1-LOW + MID + Planck. For a best-fit value of $M_\nu = 0.06$ eV (i.e. at the minimum bound), this final figure would be sufficient for a $\gtrsim \!\! 2\sigma$ detection of non-zero $M_\nu$ from cosmological data alone. This would be a significant improvement over the 95\% upper limit of $M_\nu < 0.17$ eV from the Planck 2015 release, which combines CMB temperature and polarisation data with a compilation of BAO constraints \citep{Planck_2015},\footnote{These are the \texttt{Planck TT,TE,EE+lowP+BAO} constraints; see Eq.~(54d) of \cite{Planck_2015}.} and the best current upper limit of $M_\nu < 0.12$ eV, from Planck CMB + BAO + Ly$\alpha$ data \citep{ades}. Note that an upper limit of $M_\nu<0.095$ eV would also be enough to rule out the inverted hierarchy \citep{LesgourguesPastor}.

It has been shown before that various large-scale structure surveys, including IM surveys, can provide strong constraints on $M_\nu$ \citep[e.g.][]{2008PhRvD..78f5009P, 2008PhRvD..78b3529M, Carbone:2010ik, 2011APh....35..177A, Audren_2013, Font_2014, AASKA_EORCosmo, Sartoris_2015}. For example, \cite{2008PhRvD..78f5009P} found a similar constraint, $\sigma(M_\nu)=0.075$ eV at $1\sigma$, when forecasting for Planck + SKA at $z \gtrsim 8$ (during the epoch of reionization). \cite{Font_2014} forecast a significantly stronger constraint from Planck + DESI at lower redshift of $\sigma(M_\nu)=0.024$ eV, improving to $0.011$ eV when Euclid, LSST, and Lyman-$\alpha$ forest data are also included. These studies all used different intrumental and modelling assumptions however, and marginalized over different cosmological and nuisance parameters, so a direct comparison is not possible. To put our result into context, we therefore also produced Fisher forecasts for the spectroscopic survey with $M_\nu$ now included as a parameter, finding $\sigma(M_\nu) = 0.060$ eV for the combination of spectro-z + Planck $M_\nu$. Up to the same forecasting assumptions, the combination of IM surveys here should therefore be competitive with future spectroscopic surveys like Euclid and DESI in terms of a neutrino mass measurement.

Perhaps more significant is the forecast of $\sigma(M_\nu) = 0.089$ eV for SKA1-LOW + Planck. While not quite enough for a $2\sigma$ detection of $M_\nu = 0.06$ eV, it is nevertheless a strong constraint, using a completely different redshift range and observation technique to other planned surveys. This is despite our pessimistic analysis, which marginalises over the amplitude and scale-dependence of the bias in each redshift bin, and uses only the monopole of the redshift-space power spectrum. The redshift range we assumed for LOW ($z \sim 3 - 6$) has several advantages for matter power spectrum measurements -- non-linear effects are important only on significantly smaller scales than at lower redshifts, and radiative transfer processes that affect the 21cm signal at higher redshift, in the EoR \citep{AASKA_EORmodel}, are not present. An IM survey, piggy-backed on the deep EoR survey that will be performed by SKA1-LOW anyway, may therefore be an interesting prospect for a more `systematic-tolerant' neutrino mass measurement survey \citep[although foreground contamination and instrumental systematics are still serious issues that would need to be resolved; see e.g.][]{Alonso:2014dhk, AASKA_EORfg, AASKA_IMfg}.

\begin{figure}[t]
\includegraphics[width=\columnwidth]{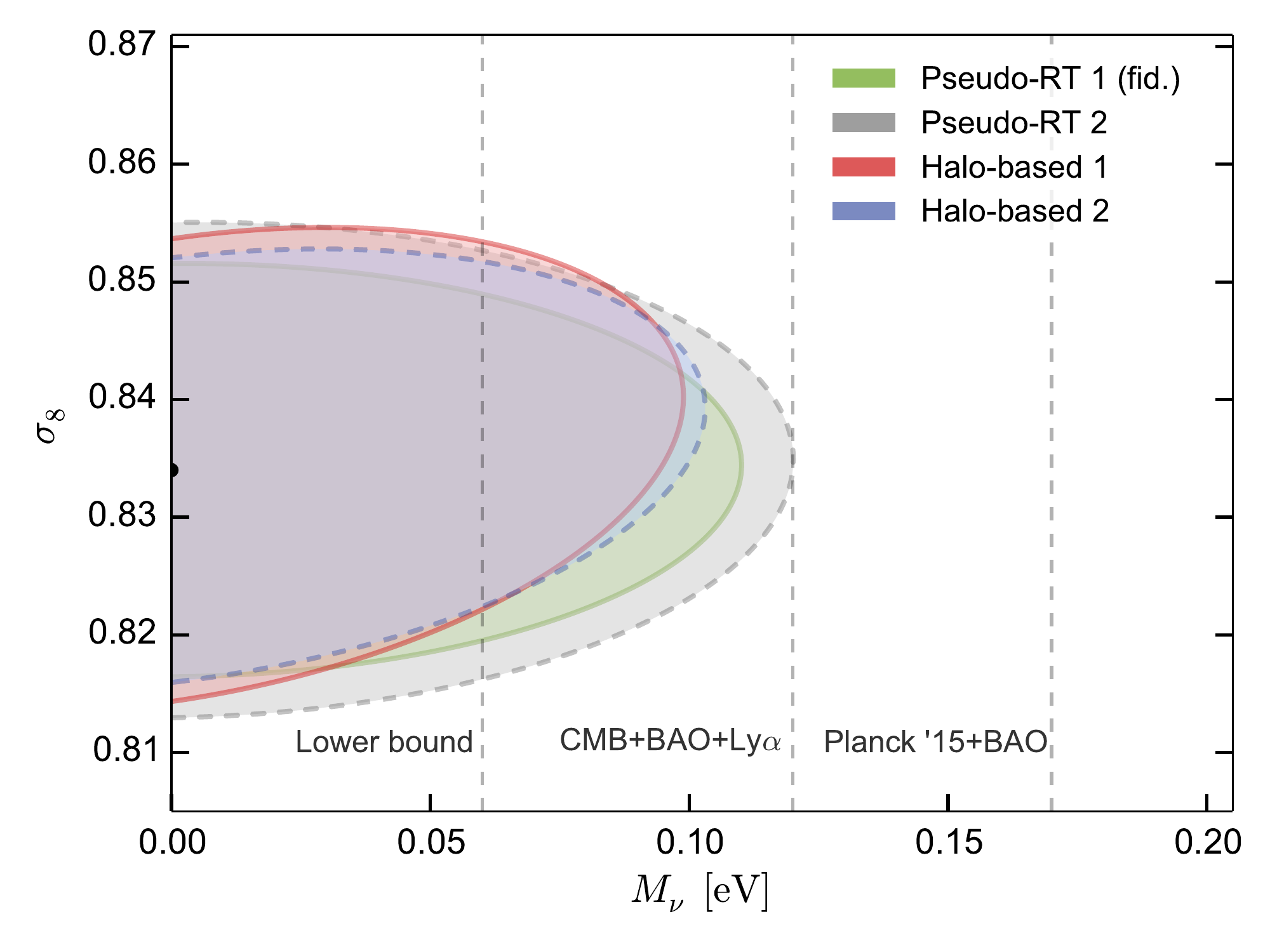}
\caption{Forecast 2$\sigma$ constraints on $M_\nu$ and $\sigma_8$ for SKA1-LOW + Planck, for the different methods used to model the HI (see Sec.~\ref{sec:HI}). The largest ellipses are obtained by employing the pseudo-RT 1 method (green, solid line) and the pseudo-RT 2 method (grey, dashed line); the smallest are for the halo-based 1 method (red, solid) and halo-based 2 method (blue, dashed).}
\label{fig:mnu_sig8_hitype}
\end{figure}

\subsection{Dependence on HI prescription and bias}
\label{sec:hidependence}

The various methods used to model the spatial distribution of HI discussed in Section \ref{sec:HI} give somewhat different predictions for the amplitude of the 21cm power spectrum, especially at high redshift. Clearly this can impact the $M_\nu$ constraints by (e.g.) changing the SNR of the power spectrum detection, so in this section we investigate the senstivity of $\sigma(M_\nu)$ to this choice. We concentrate on the SKA1-LOW + Planck constraints, including the Planck $M_\nu$ information. There is a spread of $\sigma(M_\nu)$ values (at 95\% CL), ranging from 0.089 eV for the default pseudo-RT 1 method, to 0.097 eV (pseudo-RT 2), 0.080 eV (halo-based 1), and 0.083 eV (halo-based 2); corresponding forecasts for $M_\nu$ vs. $\sigma_8$ for each HI prescription are shown in Fig.~\ref{fig:mnu_sig8_hitype}.

At low redshift, uncertainty in the bias amplitude parameters is a potential limiting factor for the neutrino mass constraint. Without any bias priors, we find that SKA1-MID (plus the Planck prior) should achieve $\sigma(M_\nu) = 0.071$ eV. A 1\% prior on the bias in all redshift bins improves this slightly to $\sigma(M_\nu) = 0.066$ eV, while a 0.1\% prior further reduces it to $0.058$ eV. A high-precision bias model is therefore of some use in improving the neutrino mass constraints at these redshifts, although the gains are too small to justify the difficulty of reaching this level of accuracy in practise. At high redshifts, the constraints are also relatively insensitive to the bias priors -- for the combination SKA1-LOW + Planck, $\sigma(M_\nu)$ improves to $0.082$ eV when a 1\% prior is applied to the $A_i$ parameters. An additional 1\% prior on the $\alpha_i$ parameters yields $\sigma(M_\nu) = 0.079$ eV.

The low redshift forecasts use 21cm power spectra calculated from the \cite{Bagla_2010} HI model for the function $M_{\rm HI}(M,z)$, which has two free parameters -- $v_{\rm min}$ and $v_{\rm max}$ -- as described in Section \ref{subsec:low-z}. These effectively set the normalisation of the 21cm power spectrum, and are constrained by the need to reproduce the observed HI density evolution. They are nevertheless subject to some uncertainty, so we also check the effect of allowing them to be free parameters. We find that there are strong correlations between $v_{\rm min}$ and $v_{\rm max}$ and the bias parameters, to the point where a weak prior on one of the two (e.g. $\sigma(v_{\rm max}) \sim 100$ kms$^{-1}$) is needed to make the Fisher matrix invertible for SKA1-MID alone. Once this has been applied, however, there is essentially no correlation between the Bagla parameters and $M_\nu$, meaning that their effect on $\sigma(M_\nu)$ is limited to changing the signal-to-noise ratio of the 21cm power spectrum detection.

\subsection{Dependence on survey parameters}
\label{sec:surveydependence}

\begin{figure}[t]
\includegraphics[width=\columnwidth]{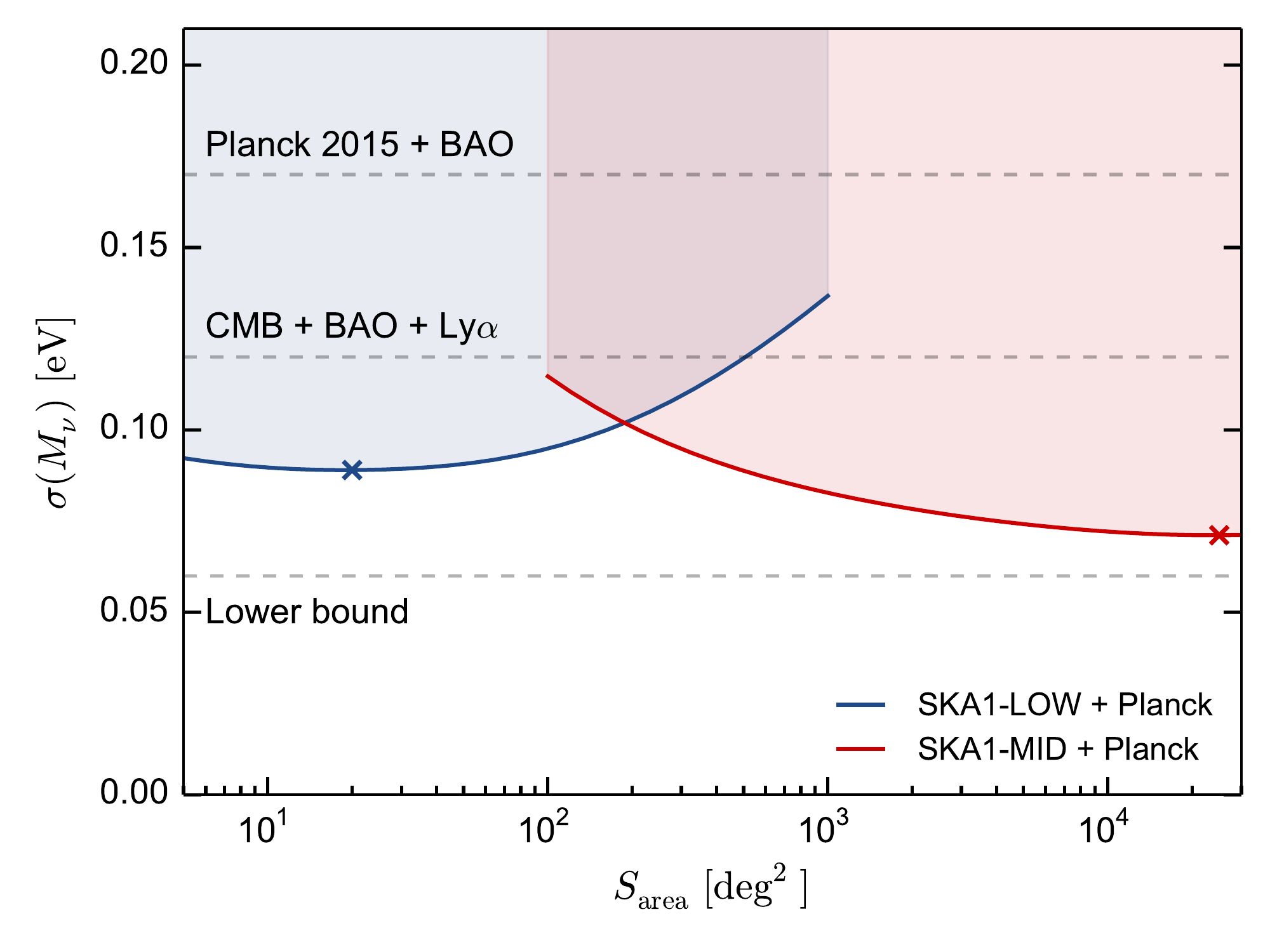}
\caption{Forecast 2$\sigma$ constraints on $M_\nu$ as a function of survey area, for the SKA1-LOW and SKA1-MID surveys combined with Planck. The current best upper and lower limits on the sum of the neutrino masses are shown as dashed lines. The fiducial survey areas are marked as crosses.}
\label{fig:mnu_sarea}
\end{figure}

Finally, we check the dependence of our results on the assumed IM survey parameters. Fig.~\ref{fig:mnu_sarea} shows $\sigma(M_\nu)$ as a function of the total survey area, $S_{\rm area}$, for SKA1-LOW and MID, combined with the full Planck Fisher matrix. Fig.~\ref{fig:mnu_ttot} shows the same as a function of survey time, $t_{\rm tot}$.

For a fixed survey time of $t_{\rm tot} = 10^4$ hours, the dependence on survey area is relatively mild for both surveys -- SKA1-LOW yields essentially the same constraints for survey areas up to $100$ deg$^2$. Similarly, the MID constraints improve relatively little above $S_{\rm area} \approx 2000$ deg$^2$. In both cases the fiducial survey areas are close to optimal for the chosen survey time.

Fixing the survey areas to their fiducial values, we find a greater sensitivity to the chosen $t_{\rm tot}$. Larger survey times than the fiducial value of 10,000 hours are largely impractical, but would be required for the constraints for either survey to cross the minimum mass value ($0.06$ eV) at the $2\sigma$ level -- MID would require 30,000 hours to reach this limit, while LOW would require 100,000 hours. MID provides a useful improvement over the current best constraint for a minimum survey time of $t_{\rm tot} \gtrsim 600$ hours, while LOW requires $t_{\rm tot} \gtrsim 3000$ hours. Note that all survey times should be thought of as `effective' values, as foreground cleaning and other effects will increase the noise level, requiring an increase in survey time to compensate.

\section{Summary and conclusions} 
\label{sec:conclusions}

Neutrinos are one of the most enigmatic particles in nature. The standard model of particle physics describe them as massless particles, while observations of the so-called neutrino oscillations imply that at least two of the three neutrino species must be massive. Massive neutrinos are therefore one of the clearest indications of physics beyond the standard model. As such, one of the most important currently-unanswered questions in modern physics is: what are the masses of the neutrinos?

Constraints on the neutrino masses arising from laboratory experiments are not very tight, $m(\bar{\nu}_e)\textless2.3$ eV \citep{Kraus_2005}, but are expected to shrink down to $m(\bar{\nu}_e)\textless0.2$ eV in the coming years. On the other hand, constraints obtained by using cosmological observables are very tight: $\sum_i m_{\nu_i} <0.12$ eV ($95\%$ CL) \citep{ades}. This constraint is obtained by combining data from CMB, BAO, and the Ly$\alpha$-forest \citep{palanque-data}.

\begin{figure}[t]
\includegraphics[width=\columnwidth]{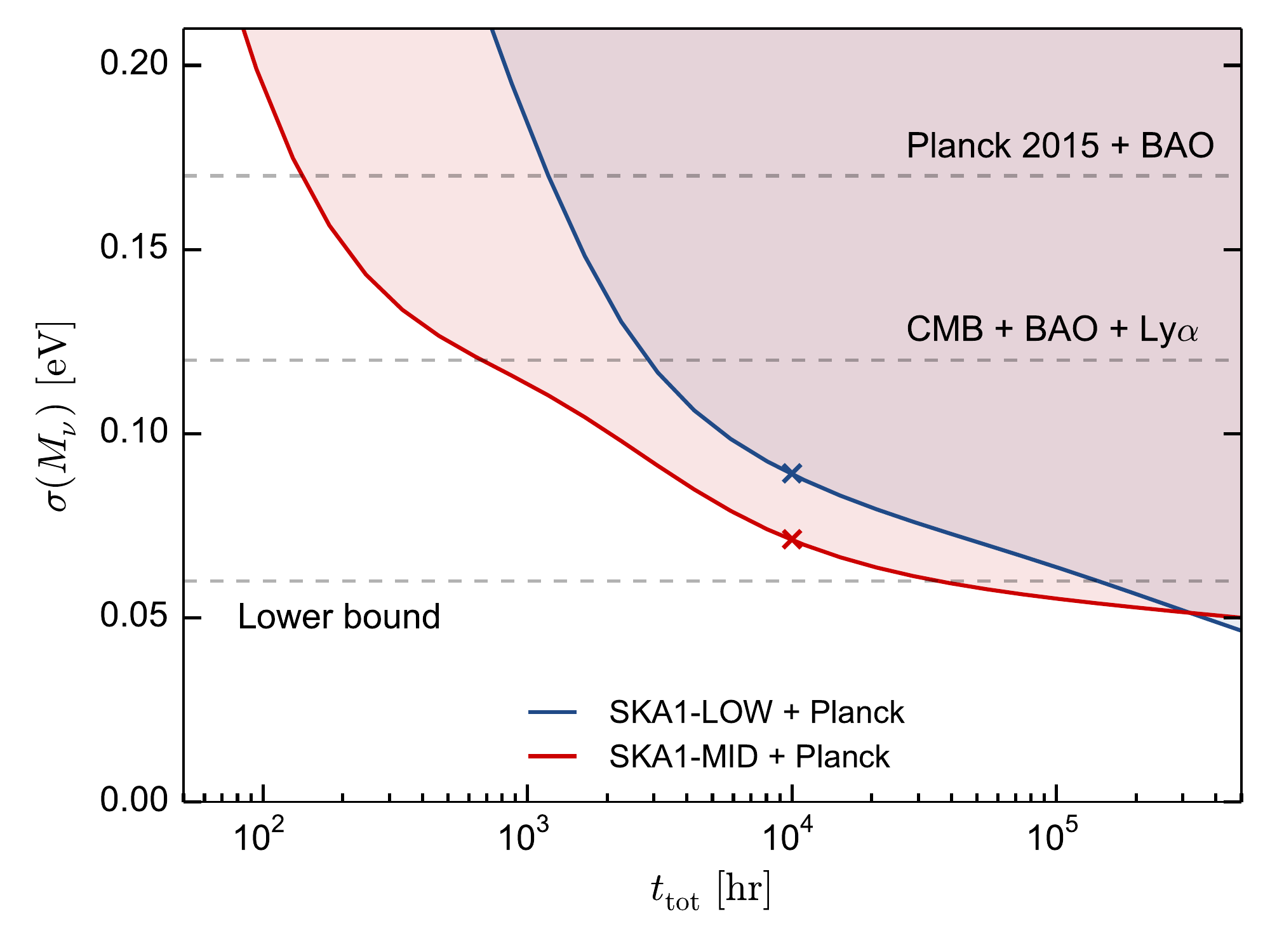}
\caption{Forecast marginal 1$\sigma$ constraint on $M_\nu$ as a function of survey time, for the same combination of surveys as in Fig.~\ref{fig:mnu_sarea}. The SKA1-LOW IM survey is taken to be 20 deg$^2$.}
\label{fig:mnu_ttot}
\end{figure}

In the near future, 21cm intensity mapping observations in the post-reionization era will be introduced as a powerful new cosmological tool \citep{Bull_2015}. These observations can also be used to place extremely tight constraints on the neutrino masses. In order to extract the maximum information from these surveys, one needs to understand, from the theory side, the impact that massive neutrinos have on the spatial distribution of neutral hydrogen, both at linear and fully non-linear level.

In this paper we study, for the first time, the detailed effects that massive neutrinos have on the neutral hydrogen spatial distribution in the post-reionization epoch, focusing our attention on their impact on the HI clustering and abundance. We do this by running hydrodynamical simulations with massless and massive neutrinos that cover the redshift range $3\leqslant z \leqslant5.5$. The output of our simulations is corrected {\it a posteriori} to account for HI self shielding and the formation of molecular hydrogen. We use four different methods to model those processes. In our fiducial method (pseudo-RT 1), the HI self-shielding is corrected by employing the fitting formula of \cite{Rahmati_2013}, while we use a phenomenological relation where the fraction of molecular hydrogen depends on the hydrodynamical pressure, based on the Blizt \& Rosolowsky and THINGS observations \citep{Blitz_2006, Leroy_2008}, to model the formation of molecular hydrogen.

We find that neutral hydrogen is more clustered in cosmologies with massive neutrinos, although its abundance, $\Omega_{\rm HI}(z)$, is lower. These differences increase with the sum of the neutrino masses, and are mainly due to the impact that massive neutrinos have on the spatial distribution of matter, which is well known (see Appendix \ref{sec:appendixA}): they suppress the amplitude of the matter power spectrum, and thus the abundance of dark matter halos, on small scales. The reason for the differences in the spatial distribution of neutral hydrogen between the massless/massive neutrino cases is mainly due to the different cosmologies; massive neutrinos barely modify the function $M_{\rm HI}(M,z)$. That is, for a fixed dark matter halo mass, the HI mass in cosmologies with massive/massless neutrinos is very similar, although on average it increases slightly with the sum of the neutrino masses. 

We find that the $M_{\rm HI}(M,z)$ function can be well fitted by $M_{\rm HI}(M,z)=(M/M_0)^\alpha$ in the redshift and mass range covered by our simulations. The value of $\alpha$ decreases with redshift and, for a fixed redshift, increases with the sum of the neutrino masses. This points out another well-known effect of massive neutrinos: in cosmologies with massive neutrinos, the spatial distribution of neutral hydrogen on small scales can be viewed as the spatial distribution of HI in the corresponding massless neutrino model at a earlier time.

In terms of the HI column density distribution function, we find that our fiducial massless neutrinos model reproduces observational measurements in the redshift range $3\leqslant z \leqslant 5$ very well. In cosmologies with massive neutrinos we find a deficit in the abundance of DLAs and sub-DLAs, which can be explained by the lower value of $\Omega_{\rm HI}(z)$ present in those cosmologies.

As stated above, the HI is more strongly clustered in cosmologies with massive neutrinos. The reason is that $M_{\rm HI}(M,z)$ barely changes, but halos of the same mass are more clustered in massive neutrino cosmologies; this happens because massive neutrinos suppress the abundance of dark matter halos, and therefore halos of the same mass will cluster more strongly in cosmologies with massive neutrinos, since they are rarer in those models.

Even though the HI is more clustered in massive neutrino cosmologies, we find that the amplitude of the 21cm power spectrum is lower in those models with respect to the model with massless neutrinos. The reason for this is again that $\Omega_{\rm HI}(z)$ is lower in those models.  

We have forecasted the constraints that the future SKA telescope will place on the sum of the neutrino masses by means of 21cm intensity mapping observations in the post-reionization era. We did this using the Fisher matrix formalism, modeling the spatial distribution of neutral hydrogen in 14 different cosmological models from redshift $z=0$ to $z \approx 6$. At redshifts $z>3$ we use hydrodynamic simulations to simulate the spatial distribution of neutral hydrogen, while at $z<3$ we use a simple analytic model based on halofit.

Our forecasts are summarised in Table \ref{tbl:marginal}. We find that with 10,000 hours of observations in a deep and narrow survey, SKA1-LOW, employing the interferometer mode, will be able to measure the sum of the neutrino masses with an error $\sigma(M_\nu)\simeq0.3$ eV (95\% CL), lower than the one obtained from CMB observations alone \citep{Planck_2015}. A slightly tighter constraint can be placed with 10,000 hours of observations with SKA1-MID, using single-dish (autocorrelation) mode for a wide 25,000 deg$^2$ survey.

Combining the SKA results with CMB data, we forecast that the sum of the neutrino masses can be constrained with a much smaller error, as various degeneracies with cosmological parameters are broken. For instance, by using CMB constraints on cosmological parameters alone (i.e. neglecting information on $M_\nu$ from the CMB), we find that SKA1-LOW+CMB will be able to measure the sum of the neutrino masses with an error $\sigma(M_\nu)\simeq0.21$ eV (95\% CL), while the combination SKA1-MID + CMB will improve this to $\sigma(M_\nu)\simeq0.19$ eV (95\% CL). On the other hand, if we do include the information on $M_\nu$ contained in the CMB, we find that $\sigma(M_\nu)\simeq0.09,0.07$ for SKA1-LOW + CMB and SKA1-MID + CMB, respectively.

Finally, by also including a prior on the cosmological parameters from a spectroscopic redshift survey such as Euclid or DESI, we see a modest further improvement. For instance, by combining data from the CMB (neglecting information on the sum of the neutrino masses), a spectro-z survey, and the SKA (either SKA1-LOW or SKA1-MID), we find that $\sigma(M_\nu)\simeq0.1$ eV. As expected, the tighter constraint on the sum of the neutrino masses can be obtained by combining the full data (including the CMB $M_\nu$ constraint), a spectro-z survey, and data from both SKA1-LOW and SKA1-MID: $\sigma(M_\nu)=0.058$ (95\% CL), which represents a significant improvement over the current tightest bound arising from CMB+BAO+Ly$\alpha$-forest data of $M_\nu<0.12$ eV \citep{ades}.

In order to check the robustness of our forecasts for SKA1-LOW, which were obtained by running high-resolution hydrodynamic simulations and modeling the HI using the pseudo-RT 1 method, we have repeated the analysis by modeling the spatial distribution of neutral hydrogen using 3 completely different methods. We find that our results are very stable against the model used to assign the HI, as can be seen in Fig.~\ref{fig:mnu_sig8_hitype}.

Our forecasts depend only weakly on the assumed survey area (see Fig.~\ref{fig:mnu_sarea}), but are more sensitive to the total observation time available (Fig.~\ref{fig:mnu_ttot}). A 10,000 hour survey performed `commensally' with a deep EoR survey over 5 pointings on the sky with SKA1-LOW is a realistic prospect however, and should be able to put strong constraints on the sum of neutrino masses.

We conclude that massive neutrinos imprint distinctive signatures on the 21cm power spectrum in the post-reionization era, and that future 21cm intensity mapping surveys with (e.g.) the SKA will be able to place very tight constraints on the sum of the neutrino masses.

\begin{acknowledgements}
We thank Mario G. Santos for useful discussions. Simulations were performed on the COSMOS Consortium supercomputer within the DiRAC Facility jointly funded by STFC, the Large Facilities Capital Fund of BIS and the University of Cambridge, as well as the Darwin Supercomputer of the University of Cambridge High Performance Computing Service (\url{http://www.hpc.cam.ac.uk/}), provided by Dell Inc. using Strategic Research Infrastructure Funding from the Higher Education Funding Council for England.  Part of the simulations and the post-processing has been carried out in the Zefiro cluster (Pisa, Italy). FVN and MV are supported by the ERC Starting Grant ``cosmoIGM'' under GA 257670 and partially supported by INFN IS PD51 ``INDARK''. PB is supported by ERC grant StG2010-257080. We acknowledge partial support from ``Consorzio per la Fisica -- Trieste''.
\end{acknowledgements}

\bibliography{papers_bibtex}{}

\begin{thebibliography}{122}
\expandafter\ifx\csname natexlab\endcsname\relax\def\natexlab#1{#1}\fi

\bibitem[{{Abazajian} {et~al.}(2011){Abazajian}, {Calabrese}, {Cooray}, {De
  Bernardis}, {Dodelson}, {Friedland}, {Fuller}, {Hannestad}, {Keating},
  {Linder}, {Lunardini}, {Melchiorri}, {Miquel}, {Pierpaoli}, {Pritchard},
  {Serra}, {Takada}, \& {Wong}}]{2011APh....35..177A}
{Abazajian}, K.~N. {et~al.} 2011, Astroparticle Physics, 35, 177,
  [arXiv:1103.5083]

\bibitem[{{Agarwal} \& {Feldman}(2011)}]{Agarwal2011}
{Agarwal}, S., \& {Feldman}, H.~A. 2011, \mnras, 410, 1647, [arXiv:1006.0689]

\bibitem[{{Ali-Ha{\"i}moud} \& {Bird}(2013)}]{Yacine-Bird}
{Ali-Ha{\"i}moud}, Y., \& {Bird}, S. 2013, \mnras, 428, 3375, [arXiv:1209.0461]

\bibitem[{Alonso {et~al.}(2015)Alonso, Bull, Ferreira, \&
  Santos}]{Alonso:2014dhk}
Alonso, D., Bull, P., Ferreira, P.~G., \& Santos, M.~G. 2015, Mon. Not. Roy.
  Astron. Soc., 447, 400, [arXiv:1409.8667]

\bibitem[{Amendola {et~al.}(2013)}]{Amendola:2012ys}
Amendola, L., {et~al.} 2013, Living Rev.Rel., 16, 6, [arXiv:1206.1225]

\bibitem[{{Audren} {et~al.}(2013){Audren}, {Lesgourgues}, {Bird}, {Haehnelt},
  \& {Viel}}]{Audren_2013}
{Audren}, B., {Lesgourgues}, J., {Bird}, S., {Haehnelt}, M.~G., \& {Viel}, M.
  2013, \jcap, 1, 26, [arXiv:1210.2194]

\bibitem[{{Bagla} {et~al.}(2010){Bagla}, {Khandai}, \& {Datta}}]{Bagla_2010}
{Bagla}, J.~S., {Khandai}, N., \& {Datta}, K.~K. 2010, \mnras, 407, 567,
  [arXiv:0908.3796]

\bibitem[{{Basse} {et~al.}(2013){Basse}, {Eggers Bjaelde}, {Hamann},
  {Hannestad}, \& {Wong}}]{Basse}
{Basse}, T., {Eggers Bjaelde}, O., {Hamann}, J., {Hannestad}, S., \& {Wong},
  Y.~Y.~Y. 2013, ArXiv e-prints, [arXiv:1304.2321]

\bibitem[{Battye {et~al.}(2004)Battye, Davies, \& Weller}]{Battye:2004re}
Battye, R.~A., Davies, R.~D., \& Weller, J. 2004, Mon.Not.Roy.Astron.Soc., 355,
  1339, [arXiv:astro-ph/0401340]

\bibitem[{{Battye} \& {Moss}(2014)}]{Battye_2013}
{Battye}, R.~A., \& {Moss}, A. 2014, Physical Review Letters, 112, 051303,
  [arXiv:1308.5870]

\bibitem[{{Beutler} {et~al.}(2014){Beutler}, {Saito}, {Brownstein}, {Chuang},
  {Cuesta}, {Percival}, {Ross}, {Ross}, {Schneider}, {Samushia}, {S{\'a}nchez},
  {Seo}, {Tinker}, {Wagner}, \& {Weaver}}]{Beutler_2014}
{Beutler}, F. {et~al.} 2014, ArXiv e-prints, [arXiv:1403.4599]

\bibitem[{{Bharadwaj} {et~al.}(2001){Bharadwaj}, {Nath}, \&
  {Sethi}}]{Bharadwaj_2001A}
{Bharadwaj}, S., {Nath}, B.~B., \& {Sethi}, S.~K. 2001, Journal of Astrophysics
  and Astronomy, 22, 21, [arXiv:astro-ph/0003200]

\bibitem[{{Bharadwaj} \& {Sethi}(2001)}]{Bharadwaj_2001B}
{Bharadwaj}, S., \& {Sethi}, S.~K. 2001, Journal of Astrophysics and Astronomy,
  22, 293, [arXiv:astro-ph/0203269]

\bibitem[{{Bird} {et~al.}(2012){Bird}, {Viel}, \& {Haehnelt}}]{Bird_2011}
{Bird}, S., {Viel}, M., \& {Haehnelt}, M.~G. 2012, \mnras, 420, 2551,
  [arXiv:1109.4416]

\bibitem[{{Blas} {et~al.}(2014){Blas}, {Garny}, {Konstandin}, \&
  {Lesgourgues}}]{Blas_2014}
{Blas}, D., {Garny}, M., {Konstandin}, T., \& {Lesgourgues}, J. 2014, \jcap,
  11, 39, [arXiv:1408.2995]

\bibitem[{{Blitz} \& {Rosolowsky}(2006)}]{Blitz_2006}
{Blitz}, L., \& {Rosolowsky}, E. 2006, \apj, 650, 933, [arXiv:astro-ph/0605035]

\bibitem[{{Brandbyge} {et~al.}(2008){Brandbyge}, {Hannestad}, {Haugb{\o}lle},
  \& {Thomsen}}]{Brandbyge_2008}
{Brandbyge}, J., {Hannestad}, S., {Haugb{\o}lle}, T., \& {Thomsen}, B. 2008,
  \jcap, 8, 20, [arXiv:0802.3700]

\bibitem[{{Brandbyge} {et~al.}(2010){Brandbyge}, {Hannestad}, {Haugb{\o}lle},
  \& {Wong}}]{Brandbyge_2010}
{Brandbyge}, J., {Hannestad}, S., {Haugb{\o}lle}, T., \& {Wong}, Y.~Y.~Y. 2010,
  \jcap, 9, 14, [arXiv:1004.4105]

\bibitem[{{Bull} {et~al.}(2015){Bull}, {Ferreira}, {Patel}, \&
  {Santos}}]{Bull_2015}
{Bull}, P., {Ferreira}, P.~G., {Patel}, P., \& {Santos}, M.~G. 2015, \apj, 803,
  21, [arXiv:1405.1452]

\bibitem[{Carbone {et~al.}(2011)Carbone, Verde, Wang, \&
  Cimatti}]{Carbone:2010ik}
Carbone, C., Verde, L., Wang, Y., \& Cimatti, A. 2011, JCAP, 1103, 030,
  [arXiv:1012.2868]

\bibitem[{{Carucci} {et~al.}(2015){Carucci}, {Villaescusa-Navarro}, {Viel}, \&
  {Lapi}}]{Carucci_2015}
{Carucci}, I.~P., {Villaescusa-Navarro}, F., {Viel}, M., \& {Lapi}, A. 2015,
  ArXiv e-prints, [arXiv:1502.06961]

\bibitem[{{Castorina} {et~al.}(2015){Castorina}, {Carbone}, {Bel}, {Sefusatti},
  \& {Dolag}}]{Castorina_2015}
{Castorina}, E., {Carbone}, C., {Bel}, J., {Sefusatti}, E., \& {Dolag}, K.
  2015, ArXiv e-prints, [arXiv:1505.07148]

\bibitem[{{Castorina} {et~al.}(2014){Castorina}, {Sefusatti}, {Sheth},
  {Villaescusa-Navarro}, \& {Viel}}]{Castorina_2014}
{Castorina}, E., {Sefusatti}, E., {Sheth}, R.~K., {Villaescusa-Navarro}, F., \&
  {Viel}, M. 2014, \jcap, 2, 49, [arXiv:1311.1212]

\bibitem[{{Chang} {et~al.}(2008){Chang}, {Pen}, {Peterson}, \&
  {McDonald}}]{Chang_2008}
{Chang}, T.-C., {Pen}, U.-L., {Peterson}, J.~B., \& {McDonald}, P. 2008,
  Physical Review Letters, 100, 091303, [arXiv:0709.3672]

\bibitem[{Chapman {et~al.}(2015)Chapman, Bonaldi, Harker,
  {et~al.}}]{AASKA_EORfg}
Chapman, E., Bonaldi, A., Harker, G., {et~al.} 2015, PoS, AASKA14, 005

\bibitem[{{Cooray} \& {Sheth}(2002)}]{Cooray_Sheth_2002}
{Cooray}, A., \& {Sheth}, R. 2002, \physrep, 372, 1, [arXiv:astro-ph/0206508]

\bibitem[{{Costanzi} {et~al.}(2014){Costanzi}, {Sartoris}, {Viel}, \&
  {Borgani}}]{Costanzi_2014b}
{Costanzi}, M., {Sartoris}, B., {Viel}, M., \& {Borgani}, S. 2014, ArXiv
  e-prints, [arXiv:1407.8338]

\bibitem[{{Costanzi} {et~al.}(2013){Costanzi}, {Villaescusa-Navarro}, {Viel},
  {Xia}, {Borgani}, {Castorina}, \& {Sefusatti}}]{Costanzi_2014}
{Costanzi}, M., {Villaescusa-Navarro}, F., {Viel}, M., {Xia}, J.-Q., {Borgani},
  S., {Castorina}, E., \& {Sefusatti}, E. 2013, \jcap, 12, 12,
  [arXiv:1311.1514]

\bibitem[{{Costanzi Alunno Cerbolini} {et~al.}(2013){Costanzi Alunno
  Cerbolini}, {Sartoris}, {Xia}, {Biviano}, {Borgani}, \& {Viel}}]{Costanzi}
{Costanzi Alunno Cerbolini}, M., {Sartoris}, B., {Xia}, J.-Q., {Biviano}, A.,
  {Borgani}, S., \& {Viel}, M. 2013, \jcap, 6, 20, [arXiv:1303.4550]

\bibitem[{{Crighton} {et~al.}(2015){Crighton}, {Murphy}, {Prochaska},
  {Worseck}, {Rafelski}, {Becker}, {Ellison}, {Fumagalli}, {Lopez}, {Meiksin},
  \& {O'Meara}}]{Crighton_2015}
{Crighton}, N.~H.~M. {et~al.} 2015, ArXiv e-prints, [arXiv:1506.02037]

\bibitem[{{Dav{\'e}} {et~al.}(2013){Dav{\'e}}, {Katz}, {Oppenheimer},
  {Kollmeier}, \& {Weinberg}}]{Dave_2013}
{Dav{\'e}}, R., {Katz}, N., {Oppenheimer}, B.~D., {Kollmeier}, J.~A., \&
  {Weinberg}, D.~H. 2013, \mnras, 434, 2645, [arXiv:1302.3631]

\bibitem[{{Davis} {et~al.}(1985){Davis}, {Efstathiou}, {Frenk}, \&
  {White}}]{FoF}
{Davis}, M., {Efstathiou}, G., {Frenk}, C.~S., \& {White}, S.~D.~M. 1985, \apj,
  292, 371

\bibitem[{{de Putter} {et~al.}(2012){de Putter}, {Mena}, {Giusarma}, {Ho},
  {Cuesta}, {Seo}, {Ross}, {White}, {Bizyaev}, {Brewington}, {Kirkby},
  {Malanushenko}, {Malanushenko}, {Oravetz}, {Pan}, {Percival}, {Ross},
  {Schneider}, {Shelden}, {Simmons}, \& {Snedden}}]{dePutter}
{de Putter}, R. {et~al.} 2012, \apj, 761, 12, [arXiv:1201.1909]

\bibitem[{{Dell'Oro} {et~al.}(2015){Dell'Oro}, {Marcocci}, {Viel}, \&
  {Vissani}}]{delloro}
{Dell'Oro}, S., {Marcocci}, S., {Viel}, M., \& {Vissani}, F. 2015, ArXiv
  e-prints, [arXiv:1505.02722]

\bibitem[{Dewdney {et~al.}(2013)}]{dewdney2013ska1}
Dewdney, P., {et~al.} 2013,
  \url{https://www.skatelescope.org/wp-content/uploads/2013/03/SKA-TEL-SKO-DD-001-1\_BaselineDesign1.pdf}

\bibitem[{{Dolag} {et~al.}(2009){Dolag}, {Borgani}, {Murante}, \&
  {Springel}}]{Dolag_2009}
{Dolag}, K., {Borgani}, S., {Murante}, G., \& {Springel}, V. 2009, \mnras, 399,
  497, [arXiv:0808.3401]

\bibitem[{{Faucher-Gigu{\`e}re} {et~al.}(2010){Faucher-Gigu{\`e}re}, {Kere{\v
  s}}, {Dijkstra}, {Hernquist}, \& {Zaldarriaga}}]{Faucher-Giguere_2010}
{Faucher-Gigu{\`e}re}, C.-A., {Kere{\v s}}, D., {Dijkstra}, M., {Hernquist},
  L., \& {Zaldarriaga}, M. 2010, \apj, 725, 633, [arXiv:1005.3041]

\bibitem[{{Fogli} {et~al.}(2012){Fogli}, {Lisi}, {Marrone}, {Montanino},
  {Palazzo}, \& {Rotunno}}]{Fogli}
{Fogli}, G.~L., {Lisi}, E., {Marrone}, A., {Montanino}, D., {Palazzo}, A., \&
  {Rotunno}, A.~M. 2012, \prd, 86, 013012, [arXiv:1205.5254]

\bibitem[{{Font-Ribera} {et~al.}(2014){Font-Ribera}, {McDonald}, {Mostek},
  {Reid}, {Seo}, \& {Slosar}}]{Font_2014}
{Font-Ribera}, A., {McDonald}, P., {Mostek}, N., {Reid}, B.~A., {Seo}, H.-J.,
  \& {Slosar}, A. 2014, \jcap, 5, 23, [arXiv:1308.4164]

\bibitem[{{Font-Ribera} {et~al.}(2012){Font-Ribera}, {Miralda-Escud{\'e}},
  {Arnau}, {Carithers}, {Lee}, {Noterdaeme}, {P{\^a}ris}, {Petitjean}, {Rich},
  {Rollinde}, {Ross}, {Schneider}, {White}, \& {York}}]{Font_2012}
{Font-Ribera}, A. {et~al.} 2012, \jcap, 11, 59, [arXiv:1209.4596]

\bibitem[{{Forero} {et~al.}(2012){Forero}, {T{\'o}rtola}, \& {Valle}}]{Tortola}
{Forero}, D.~V., {T{\'o}rtola}, M., \& {Valle}, J.~W.~F. 2012, \prd, 86,
  073012, [arXiv:1205.4018]

\bibitem[{{F{\"u}hrer} \& {Wong}(2015)}]{2015JCAP...03..046F}
{F{\"u}hrer}, F., \& {Wong}, Y.~Y.~Y. 2015, \jcap, 3, 46, [arXiv:1412.2764]

\bibitem[{{Giusarma} {et~al.}(2014){Giusarma}, {Di Valentino}, {Lattanzi},
  {Melchiorri}, \& {Mena}}]{Giusarma_2014}
{Giusarma}, E., {Di Valentino}, E., {Lattanzi}, M., {Melchiorri}, A., \&
  {Mena}, O. 2014, ArXiv e-prints, [arXiv:1403.4852]

\bibitem[{{Hamann} \& {Hasenkamp}(2013)}]{Hamann_2013}
{Hamann}, J., \& {Hasenkamp}, J. 2013, \jcap, 10, 44, [arXiv:1308.3255]

\bibitem[{{Hannestad}(2003)}]{Hannestad_2003}
{Hannestad}, S. 2003, \jcap, 5, 4, [arXiv:arXiv:astro-ph/0303076]

\bibitem[{{Haynes} {et~al.}(2011){Haynes}, {Giovanelli}, {Martin}, {Hess},
  {Saintonge}, {Adams}, {Hallenbeck}, {Hoffman}, {Huang}, {Kent}, {Koopmann},
  {Papastergis}, {Stierwalt}, {Balonek}, {Craig}, {Higdon}, {Kornreich},
  {Miller}, {O'Donoghue}, {Olowin}, {Rosenberg}, {Spekkens}, {Troischt}, \&
  {Wilcots}}]{Haynes_2011}
{Haynes}, M.~P. {et~al.} 2011, \aj, 142, 170, [arXiv:1109.0027]

\bibitem[{{Ichiki} \& {Takada}(2012)}]{Ichiki-Takada}
{Ichiki}, K., \& {Takada}, M. 2012, \prd, 85, 063521, [arXiv:1108.4688]

\bibitem[{{Inman} {et~al.}(2015){Inman}, {Emberson}, {Pen}, {Farchi}, {Yu}, \&
  {Harnois-Deraps}}]{Inman_2015}
{Inman}, D., {Emberson}, J.~D., {Pen}, U.-L., {Farchi}, A., {Yu}, H.-R., \&
  {Harnois-Deraps}, J. 2015, ArXiv e-prints, [arXiv:1503.07480]

\bibitem[{{Kaiser}(1987)}]{Kaiser_1987}
{Kaiser}, N. 1987, \mnras, 227, 1

\bibitem[{{Kaplinghat} {et~al.}(2003){Kaplinghat}, {Knox}, \&
  {Song}}]{2003PhRvL..91x1301K}
{Kaplinghat}, M., {Knox}, L., \& {Song}, Y.-S. 2003, Physical Review Letters,
  91, 241301, [arXiv:astro-ph/0303344]

\bibitem[{Kitching {et~al.}(2008)Kitching, Heavens, Verde, Serra, \&
  Melchiorri}]{Kitching:2008dp}
Kitching, T., Heavens, A., Verde, L., Serra, P., \& Melchiorri, A. 2008,
  Phys.Rev., D77, 103008, [arXiv:0801.4565]

\bibitem[{Koopmans {et~al.}(2015)Koopmans, Pritchard, Mellema,
  {et~al.}}]{AASKA_EOR}
Koopmans, L.~V.~E., Pritchard, J., Mellema, G., {et~al.} 2015, PoS, AASKA14,
  001

\bibitem[{{Kraus} {et~al.}(2005){Kraus}, {Bornschein}, {Bornschein}, {Bonn},
  {Flatt}, {Kovalik}, {Ostrick}, {Otten}, {Schall}, {Th{\"u}mmler}, \&
  {Weinheimer}}]{Kraus_2005}
{Kraus}, C. {et~al.} 2005, European Physical Journal C, 40, 447,
  [arXiv:hep-ex/0412056]

\bibitem[{{Leroy} {et~al.}(2008){Leroy}, {Walter}, {Brinks}, {Bigiel}, {de
  Blok}, {Madore}, \& {Thornley}}]{Leroy_2008}
{Leroy}, A.~K., {Walter}, F., {Brinks}, E., {Bigiel}, F., {de Blok}, W.~J.~G.,
  {Madore}, B., \& {Thornley}, M.~D. 2008, \aj, 136, 2782, [arXiv:0810.2556]

\bibitem[{{Lesgourgues} {et~al.}(2013){Lesgourgues}, {Mangano}, {Miele}, \&
  {Pastor}}]{LesgourguesBook}
{Lesgourgues}, J., {Mangano}, G., {Miele}, G., \& {Pastor}. 2013, {Neutrino
  cosmology}

\bibitem[{{Lesgourgues} \& {Pastor}(2006)}]{LesgourguesPastor}
{Lesgourgues}, J., \& {Pastor}, S. 2006, \physrep, 429, 307,
  [arXiv:arXiv:astro-ph/0603494]

\bibitem[{{Lewis} {et~al.}(2000){Lewis}, {Challinor}, \& {Lasenby}}]{CAMB}
{Lewis}, A., {Challinor}, A., \& {Lasenby}, A. 2000, \apj, 538, 473,
  [arXiv:arXiv:astro-ph/9911177]

\bibitem[{{Loeb} \& {Wyithe}(2008)}]{Loeb_Wyithe_2008}
{Loeb}, A., \& {Wyithe}, J.~S.~B. 2008, Physical Review Letters, 100, 161301,
  [arXiv:0801.1677]

\bibitem[{{LoVerde}(2014)}]{LoVerde_2014}
{LoVerde}, M. 2014, \prd, 90, 083518, [arXiv:1405.4858]

\bibitem[{{Ma} \& {Bertschinger}(1994)}]{1994ApJ...434L...5M}
{Ma}, C.-P., \& {Bertschinger}, E. 1994, \apjl, 434, L5,
  [arXiv:astro-ph/9407085]

\bibitem[{{Ma} \& {Bertschinger}(1995)}]{1995ApJ...455....7M}
------. 1995, \apj, 455, 7, [arXiv:astro-ph/9506072]

\bibitem[{{Mao} {et~al.}(2008){Mao}, {Tegmark}, {McQuinn}, {Zaldarriaga}, \&
  {Zahn}}]{2008PhRvD..78b3529M}
{Mao}, Y., {Tegmark}, M., {McQuinn}, M., {Zaldarriaga}, M., \& {Zahn}, O. 2008,
  \prd, 78, 023529, [arXiv:0802.1710]

\bibitem[{{Mar{\'{\i}}n} {et~al.}(2010){Mar{\'{\i}}n}, {Gnedin}, {Seo}, \&
  {Vallinotto}}]{Marin_2010}
{Mar{\'{\i}}n}, F.~A., {Gnedin}, N.~Y., {Seo}, H.-J., \& {Vallinotto}, A. 2010,
  \apj, 718, 972, [arXiv:0911.0041]

\bibitem[{{Marulli} {et~al.}(2011){Marulli}, {Carbone}, {Viel}, {Moscardini},
  \& {Cimatti}}]{Marulli_2011}
{Marulli}, F., {Carbone}, C., {Viel}, M., {Moscardini}, L., \& {Cimatti}, A.
  2011, \mnras, 418, 346, [arXiv:1103.0278]

\bibitem[{{Massara} {et~al.}(2014){Massara}, {Villaescusa-Navarro}, \&
  {Viel}}]{Massara_2014}
{Massara}, E., {Villaescusa-Navarro}, F., \& {Viel}, M. 2014, \jcap, 12, 53,
  [arXiv:1410.6813]

\bibitem[{{Massara} {et~al.}(2015){Massara}, {Villaescusa-Navarro}, {Viel}, \&
  {Sutter}}]{Massara_2015}
{Massara}, E., {Villaescusa-Navarro}, F., {Viel}, M., \& {Sutter}, P.~M. 2015,
  ArXiv e-prints, [arXiv:1506.03088]

\bibitem[{{McQuinn} {et~al.}(2006){McQuinn}, {Zahn}, {Zaldarriaga},
  {Hernquist}, \& {Furlanetto}}]{McQuinn_2006}
{McQuinn}, M., {Zahn}, O., {Zaldarriaga}, M., {Hernquist}, L., \& {Furlanetto},
  S.~R. 2006, \apj, 653, 815, [arXiv:astro-ph/0512263]

\bibitem[{{Metcalf}(2010)}]{Metcalf_2010}
{Metcalf}, R.~B. 2010, \mnras, 401, 1999, [arXiv:0901.0245]

\bibitem[{{Miralda-Escud{\'e}}(2005)}]{Miralda-Escude_2005}
{Miralda-Escud{\'e}}, J. 2005, \apjl, 620, L91, [arXiv:astro-ph/0410315]

\bibitem[{{Mitra} {et~al.}(2015){Mitra}, {Choudhury}, \&
  {Ferrara}}]{Mitra_2015}
{Mitra}, S., {Choudhury}, T.~R., \& {Ferrara}, A. 2015, \mnras, 454, L76,
  [arXiv:1505.05507]

\bibitem[{{Noterdaeme} {et~al.}(2012){Noterdaeme}, {Petitjean}, {Carithers},
  {P{\^a}ris}, {Font-Ribera}, {Bailey}, {Aubourg}, {Bizyaev}, {Ebelke},
  {Finley}, {Ge}, {Malanushenko}, {Malanushenko}, {Miralda-Escud{\'e}},
  {Myers}, {Oravetz}, {Pan}, {Pieri}, {Ross}, {Schneider}, {Simmons}, \&
  {York}}]{Noterdaeme_2012}
{Noterdaeme}, P. {et~al.} 2012, \aap, 547, L1, [arXiv:1210.1213]

\bibitem[{{Oyama} {et~al.}(2013){Oyama}, {Shimizu}, \& {Kohri}}]{Oyama_2012}
{Oyama}, Y., {Shimizu}, A., \& {Kohri}, K. 2013, Physics Letters B, 718, 1186,
  [arXiv:1205.5223]

\bibitem[{{Padmanabhan} {et~al.}(2015){Padmanabhan}, {Choudhury}, \&
  {Refregier}}]{Padmanabhan_2015}
{Padmanabhan}, H., {Choudhury}, T.~R., \& {Refregier}, A. 2015, ArXiv e-prints,
  [arXiv:1505.00008]

\bibitem[{{Palanque-Delabrouille}
  {et~al.}(2015{\natexlab{a}}){Palanque-Delabrouille}, {Yeche}, {Baur},
  {Magneville}, {Rossi}, {Lesgourgues}, {Borde}, {Burtin}, {LeGoff}, {Rich},
  {Viel}, \& {Weinberg}}]{ades}
{Palanque-Delabrouille}, N. {et~al.} 2015{\natexlab{a}}, ArXiv e-prints,
  [arXiv:1506.05976]

\bibitem[{{Palanque-Delabrouille} {et~al.}(2013){Palanque-Delabrouille},
  {Y{\`e}che}, {Borde}, {Le Goff}, {Rossi}, {Viel}, {Aubourg}, {Bailey},
  {Bautista}, {Blomqvist}, {Bolton}, {Bolton}, {Busca}, {Carithers}, {Croft},
  {Dawson}, {Delubac}, {Font-Ribera}, {Ho}, {Kirkby}, {Lee}, {Margala},
  {Miralda-Escud{\'e}}, {Muna}, {Myers}, {Noterdaeme}, {P{\^a}ris},
  {Petitjean}, {Pieri}, {Rich}, {Rollinde}, {Ross}, {Schlegel}, {Schneider},
  {Slosar}, \& {Weinberg}}]{palanque-data}
------. 2013, \aap, 559, A85, [arXiv:1306.5896]

\bibitem[{{Palanque-Delabrouille}
  {et~al.}(2015{\natexlab{b}}){Palanque-Delabrouille}, {Y{\`e}che},
  {Lesgourgues}, {Rossi}, {Borde}, {Viel}, {Aubourg}, {Kirkby}, {LeGoff},
  {Rich}, {Roe}, {Ross}, {Schneider}, \&
  {Weinberg}}]{Palanque-Delabrouille:2014jca}
------. 2015{\natexlab{b}}, \jcap, 2, 45, [arXiv:1410.7244]

\bibitem[{{Peloso} {et~al.}(2015){Peloso}, {Pietroni}, {Viel}, \&
  {Villaescusa-Navarro}}]{BAO_neutrinos}
{Peloso}, M., {Pietroni}, M., {Viel}, M., \& {Villaescusa-Navarro}, F. 2015,
  \jcap, 7, 1, [arXiv:1505.07477]

\bibitem[{Perotto {et~al.}(2006)Perotto, Lesgourgues, Hannestad, Tu, \&
  Wong}]{Perotto:2006rj}
Perotto, L., Lesgourgues, J., Hannestad, S., Tu, H., \& Wong, Y.~Y. 2006, JCAP,
  0610, 013, [arXiv:astro-ph/0606227]

\bibitem[{{Planck Collaboration}(2013)}]{Planck_2013}
{Planck Collaboration}. 2013, ArXiv e-prints, [arXiv:1303.5076]

\bibitem[{{Planck Collaboration}(2015)}]{Planck_2015}
------. 2015, ArXiv e-prints, [arXiv:1502.01589]

\bibitem[{Pritchard {et~al.}(2015)Pritchard, Ichiki, Mesinger,
  {et~al.}}]{AASKA_EORCosmo}
Pritchard, J., Ichiki, K., Mesinger, A., {et~al.} 2015, PoS, AASKA14, 012

\bibitem[{{Pritchard} \& {Pierpaoli}(2008)}]{2008PhRvD..78f5009P}
{Pritchard}, J.~R., \& {Pierpaoli}, E. 2008, \prd, 78, 065009,
  [arXiv:0805.1920]

\bibitem[{{Rahmati} {et~al.}(2013{\natexlab{a}}){Rahmati}, {Pawlik},
  {Raicevic}, \& {Schaye}}]{Rahmati_2013}
{Rahmati}, A., {Pawlik}, A.~H., {Raicevic}, M., \& {Schaye}, J.
  2013{\natexlab{a}}, \mnras, 430, 2427, [arXiv:1210.7808]

\bibitem[{{Rahmati} {et~al.}(2013{\natexlab{b}}){Rahmati}, {Schaye}, {Pawlik},
  \& {Rai$\check{{\rm c}}$evi$\grave{{\rm c}}$}}]{Rahmati_sources}
{Rahmati}, A., {Schaye}, J., {Pawlik}, A.~H., \& {Rai$\check{{\rm
  c}}$evi$\grave{{\rm c}}$}, M. 2013{\natexlab{b}}, \mnras, 431, 2261,
  [arXiv:1301.1978]

\bibitem[{{Reid} {et~al.}(2010){Reid}, {Verde}, {Jimenez}, \& {Mena}}]{Reid}
{Reid}, B.~A., {Verde}, L., {Jimenez}, R., \& {Mena}, O. 2010, \jcap, 1, 3,
  [arXiv:0910.0008]

\bibitem[{{Riemer-S{\o}rensen} {et~al.}(2012){Riemer-S{\o}rensen}, {Blake},
  {Parkinson}, {Davis}, {Brough}, {Colless}, {Contreras}, {Couch}, {Croom},
  {Croton}, {Drinkwater}, {Forster}, {Gilbank}, {Gladders}, {Glazebrook},
  {Jelliffe}, {Jurek}, {Li}, {Madore}, {Martin}, {Pimbblet}, {Poole}, {Pracy},
  {Sharp}, {Wisnioski}, {Woods}, {Wyder}, \& {Yee}}]{WiggleZ}
{Riemer-S{\o}rensen}, S. {et~al.} 2012, \prd, 85, 081101, [arXiv:1112.4940]

\bibitem[{{Roncarelli} {et~al.}(2015){Roncarelli}, {Carbone}, \&
  {Moscardini}}]{Roncarelli_2015}
{Roncarelli}, M., {Carbone}, C., \& {Moscardini}, L. 2015, \mnras, 447, 1761,
  [arXiv:1409.4285]

\bibitem[{{Rossi} {et~al.}(2014){Rossi}, {Palanque-Delabrouille}, {Borde},
  {Viel}, {Y{\`e}che}, {Bolton}, {Rich}, \& {Le Goff}}]{Rossi_2014}
{Rossi}, G., {Palanque-Delabrouille}, N., {Borde}, A., {Viel}, M., {Y{\`e}che},
  C., {Bolton}, J.~S., {Rich}, J., \& {Le Goff}, J.-M. 2014, \aap, 567, A79,
  [arXiv:1401.6464]

\bibitem[{{Saito} {et~al.}(2008){Saito}, {Takada}, \& {Taruya}}]{Saito_2008}
{Saito}, S., {Takada}, M., \& {Taruya}, A. 2008, Physical Review Letters, 100,
  191301, [arXiv:0801.0607]

\bibitem[{{Saito} {et~al.}(2009){Saito}, {Takada}, \& {Taruya}}]{Saito_2009}
------. 2009, \prd, 80, 083528, [arXiv:0907.2922]

\bibitem[{{Saito} {et~al.}(2011){Saito}, {Takada}, \& {Taruya}}]{Saito_2010}
------. 2011, \prd, 83, 043529, [arXiv:1006.4845]

\bibitem[{Santos {et~al.}(2015)Santos, Bull, Alonso, Camera, Ferreira,
  {et~al.}}]{Santos:2015bsa}
Santos, M.~G., Bull, P., Alonso, D., Camera, S., Ferreira, P., {et~al.} 2015,
  PoS, AASKA14, 019

\bibitem[{{Santos} {et~al.}(2015){Santos}, {Bull}, {Alonso}, {Camera},
  {Ferreira}, {Bernardi}, {Maartens}, {Viel}, {Villaescusa-Navarro}, {Abdalla},
  {Jarvis}, {Metcalf}, {Pourtsidou}, \& {Wolz}}]{Santos_2015}
{Santos}, M.~G. {et~al.} 2015, ArXiv e-prints, [arXiv:1501.03989]

\bibitem[{{Sartoris} {et~al.}(2015){Sartoris}, {Biviano}, {Fedeli}, {Bartlett},
  {Borgani}, {Costanzi}, {Giocoli}, {Moscardini}, {Weller}, {Ascaso},
  {Bardelli}, {Maurogordato}, \& {Viana}}]{Sartoris_2015}
{Sartoris}, B. {et~al.} 2015, ArXiv e-prints, [arXiv:1505.02165]

\bibitem[{{Schaye}(2006)}]{Schaye_2006}
{Schaye}, J. 2006, \apj, 643, 59, [arXiv:astro-ph/0409137]

\bibitem[{{Scoccimarro} {et~al.}(2001){Scoccimarro}, {Sheth}, {Hui}, \&
  {Jain}}]{Scoccimarro_2001}
{Scoccimarro}, R., {Sheth}, R.~K., {Hui}, L., \& {Jain}, B. 2001, \apj, 546,
  20, [arXiv:astro-ph/0006319]

\bibitem[{{Sefusatti} \& {Scoccimarro}(2005)}]{Sefusatti_2005}
{Sefusatti}, E., \& {Scoccimarro}, R. 2005, \prd, 71, 063001,
  [arXiv:astro-ph/0412626]

\bibitem[{Semelin \& Iliev(2015)}]{AASKA_EORmodel}
Semelin, B., \& Iliev, I. 2015, PoS, AASKA14, 013

\bibitem[{{Sheth} {et~al.}(2001){Sheth}, {Mo}, \& {Tormen}}]{SMT}
{Sheth}, R.~K., {Mo}, H.~J., \& {Tormen}, G. 2001, \mnras, 323, 1,
  [arXiv:astro-ph/9907024]

\bibitem[{{Sheth} \& {Tormen}(1999)}]{ST}
{Sheth}, R.~K., \& {Tormen}, G. 1999, \mnras, 308, 119,
  [arXiv:astro-ph/9901122]

\bibitem[{{Shimabukuro} {et~al.}(2014){Shimabukuro}, {Ichiki}, {Inoue}, \&
  {Yokoyama}}]{Shimabukuro_2014}
{Shimabukuro}, H., {Ichiki}, K., {Inoue}, S., \& {Yokoyama}, S. 2014, \prd, 90,
  083003, [arXiv:1403.1605]

\bibitem[{{Songaila} \& {Cowie}(2010)}]{Songaila_2010}
{Songaila}, A., \& {Cowie}, L.~L. 2010, \apj, 721, 1448, [arXiv:1007.3262]

\bibitem[{{Springel}(2005)}]{Springel_2005}
{Springel}, V. 2005, \mnras, 364, 1105, [arXiv:arXiv:astro-ph/0505010]

\bibitem[{{Springel} \& {Hernquist}(2003)}]{Springel-Hernquist_2003}
{Springel}, V., \& {Hernquist}, L. 2003, \mnras, 339, 289,
  [arXiv:astro-ph/0206393]

\bibitem[{{Springel} {et~al.}(2001){Springel}, {White}, {Tormen}, \&
  {Kauffmann}}]{Subfind}
{Springel}, V., {White}, S.~D.~M., {Tormen}, G., \& {Kauffmann}, G. 2001,
  \mnras, 328, 726, [arXiv:arXiv:astro-ph/0012055]

\bibitem[{{Swanson} {et~al.}(2010){Swanson}, {Percival}, \& {Lahav}}]{Swanson}
{Swanson}, M.~E.~C., {Percival}, W.~J., \& {Lahav}, O. 2010, \mnras, 409, 1100,
  [arXiv:1006.2825]

\bibitem[{{Takahashi} {et~al.}(2012){Takahashi}, {Sato}, {Nishimichi},
  {Taruya}, \& {Oguri}}]{Halofit_2012}
{Takahashi}, R., {Sato}, M., {Nishimichi}, T., {Taruya}, A., \& {Oguri}, M.
  2012, \apj, 761, 152, [arXiv:1208.2701]

\bibitem[{{Tegmark} \& {Zaldarriaga}(2009)}]{Tegmark_2008}
{Tegmark}, M., \& {Zaldarriaga}, M. 2009, \prd, 79, 083530, [arXiv:0805.4414]

\bibitem[{{Thomas} {et~al.}(2010){Thomas}, {Abdalla}, \& {Lahav}}]{Thomas}
{Thomas}, S.~A., {Abdalla}, F.~B., \& {Lahav}, O. 2010, Physical Review
  Letters, 105, 031301, [arXiv:0911.5291]

\bibitem[{{Viel} {et~al.}(2013){Viel}, {Becker}, {Bolton}, \&
  {Haehnelt}}]{viel13}
{Viel}, M., {Becker}, G.~D., {Bolton}, J.~S., \& {Haehnelt}, M.~G. 2013, \prd,
  88, 043502, [arXiv:1306.2314]

\bibitem[{{Viel} {et~al.}(2010){Viel}, {Haehnelt}, \& {Springel}}]{Viel_2010}
{Viel}, M., {Haehnelt}, M.~G., \& {Springel}, V. 2010, \jcap, 6, 15,
  [arXiv:1003.2422]

\bibitem[{{Villaescusa-Navarro}
  {et~al.}(2013{\natexlab{a}}){Villaescusa-Navarro}, {Bird}, {Pe{\~n}a-Garay},
  \& {Viel}}]{Villaescusa-Navarro_2013}
{Villaescusa-Navarro}, F., {Bird}, S., {Pe{\~n}a-Garay}, C., \& {Viel}, M.
  2013{\natexlab{a}}, \jcap, 3, 19, [arXiv:1212.4855]

\bibitem[{{Villaescusa-Navarro}
  {et~al.}(2014{\natexlab{a}}){Villaescusa-Navarro}, {Marulli}, {Viel},
  {Branchini}, {Castorina}, {Sefusatti}, \& {Saito}}]{Villaescusa-Navarro_2014}
{Villaescusa-Navarro}, F., {Marulli}, F., {Viel}, M., {Branchini}, E.,
  {Castorina}, E., {Sefusatti}, E., \& {Saito}, S. 2014{\natexlab{a}}, \jcap,
  3, 11, [arXiv:1311.0866]

\bibitem[{{Villaescusa-Navarro}
  {et~al.}(2014{\natexlab{b}}){Villaescusa-Navarro}, {Viel}, {Datta}, \&
  {Choudhury}}]{Villaescusa-Navarro_2014a}
{Villaescusa-Navarro}, F., {Viel}, M., {Datta}, K.~K., \& {Choudhury}, T.~R.
  2014{\natexlab{b}}, \jcap, 9, 50, [arXiv:1405.6713]

\bibitem[{{Villaescusa-Navarro}
  {et~al.}(2013{\natexlab{b}}){Villaescusa-Navarro}, {Vogelsberger}, {Viel}, \&
  {Loeb}}]{Villaescusa-Navarro_2012}
{Villaescusa-Navarro}, F., {Vogelsberger}, M., {Viel}, M., \& {Loeb}, A.
  2013{\natexlab{b}}, \mnras, 431, 3670

\bibitem[{{Wagner} {et~al.}(2012){Wagner}, {Verde}, \& {Jimenez}}]{Wagner2012}
{Wagner}, C., {Verde}, L., \& {Jimenez}, R. 2012, \apjl, 752, L31,
  [arXiv:1203.5342]

\bibitem[{Wolz {et~al.}(2015)Wolz, Abdalla, Alonso, {et~al.}}]{AASKA_IMfg}
Wolz, L., Abdalla, F.~B., Alonso, D., {et~al.} 2015, PoS, AASKA14, 035

\bibitem[{{Wong}(2008)}]{Wong_2008}
{Wong}, Y.~Y.~Y. 2008, \jcap, 10, 35, [arXiv:0809.0693]

\bibitem[{{Wyman} {et~al.}(2014){Wyman}, {Rudd}, {Vanderveld}, \&
  {Hu}}]{Wyman_2013}
{Wyman}, M., {Rudd}, D.~H., {Vanderveld}, R.~A., \& {Hu}, W. 2014, Physical
  Review Letters, 112, 051302, [arXiv:1307.7715]

\bibitem[{{Xia} {et~al.}(2012){Xia}, {Granett}, {Viel}, {Bird}, {Guzzo},
  {Haehnelt}, {Coupon}, {McCracken}, \& {Mellier}}]{Xia2012}
{Xia}, J.-Q. {et~al.} 2012, \jcap, 6, 10, [arXiv:1203.5105]

\bibitem[{{Zafar} {et~al.}(2013){Zafar}, {P{\'e}roux}, {Popping}, {Milliard},
  {Deharveng}, \& {Frank}}]{Zafar_2013}
{Zafar}, T., {P{\'e}roux}, C., {Popping}, A., {Milliard}, B., {Deharveng},
  J.-M., \& {Frank}, S. 2013, \aap, 556, A141, [arXiv:1307.0602]

\bibitem[{{Zhao} {et~al.}(2012){Zhao}, {Saito}, {Percival}, {Ross},
  {Montesano}, {Viel}, {Schneider}, {Ernst}, {Manera}, {Miralda-Escude},
  {Ross}, {Samushia}, {Sanchez}, {Swanson}, {Thomas}, {Tojeiro}, {Yeche}, \&
  {York}}]{Zhao2012}
{Zhao}, G.-B. {et~al.} 2012, ArXiv e-prints, [arXiv:1211.3741]

\end{thebibliography}
\bibliographystyle{hapj}

\appendix

\section{Impact of neutrinos on the total matter distribution} 
\label{sec:appendixA}

Here we briefly discuss the impact of massive neutrinos on the total matter spatial distribution (see Fig.~\ref{fig:Image} for a visual comparison of the spatial distribution of total matter, i.e. CDM+baryons+neutrinos+stars, between cosmologies with massless/massive neutrinos). We note that this has already been studied in many different works \citep{1994ApJ...434L...5M, 1995ApJ...455....7M, LesgourguesPastor,Saito_2008, Brandbyge_2008, Wong_2008, Saito_2009, Brandbyge_2010, Rossi_2014, LoVerde_2014,
Agarwal2011,Bird_2011,Wagner2012,Viel_2010,Villaescusa-Navarro_2012,Yacine-Bird,LesgourguesBook,Blas_2014,Massara_2014,Inman_2015,2015JCAP...03..046F, BAO_neutrinos}, but remark that the majority of N-body simulations with massive neutrinos are not hydrodynamic, with the exception of \cite{Viel_2010,Villaescusa-Navarro_2012,Rossi_2014}, and one simulation in \cite{Bird_2011}. 

\begin{figure*}[h]
\begin{center}
\includegraphics[width=0.95\textwidth]{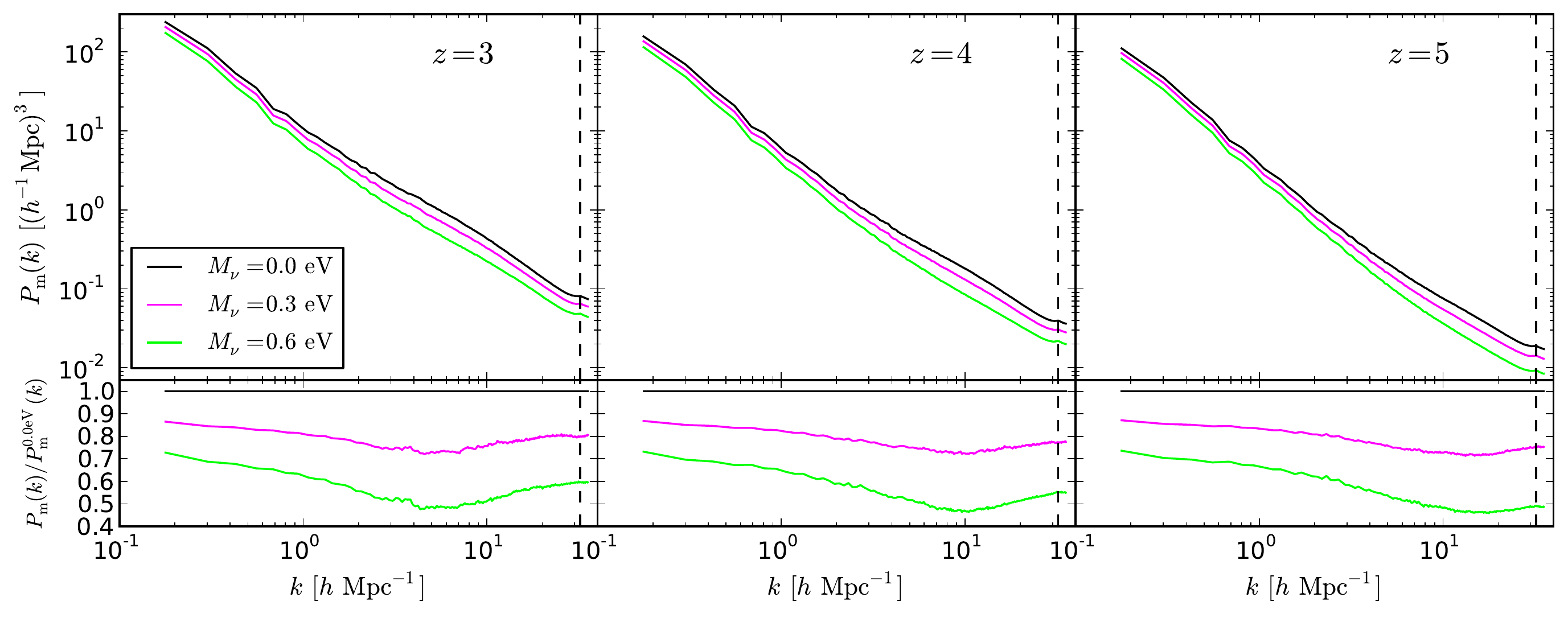}
\caption{Impact of massive neutrinos on the total matter power spectrum obtained from our high-resolution hydrodynamic simulations. Black lines show the matter power spectrum for the model with massless neutrinos at $z=3$ (left), $z=4$ (middle) and $z=5$ (right). Magenta and green lines show results for the cosmologies with $M_\nu=0.3$ eV and $M_\nu=0.6$ eV neutrinos, respectively. The bottom panels display the ratio between the matter power spectrum of the cosmologies with massive neutrinos to the massless neutrinos model. The vertical lines display the Nyquist frequency value of the grid used to measure the power spectrum.}
\label{fig:Pk_matter}
\end{center}
\end{figure*}

In Fig. \ref{fig:Pk_matter} we show the total matter power spectra for the cosmologies with $M_\nu=0.0$, 0.3 and 0.6 eV neutrinos at $z=3$, $z=4$ and $z=5$, i.e. using the simulations $\mathcal{F}$, $\nu_{\rm m}^+$ and $\nu_{\rm m}^{++}$. We find that massive neutrinos induce a suppression in the amplitude of the total matter power spectrum. The suppression increases with the sum of the neutrino masses and is almost redshift-independent. The suppression has its physical origin in the fact that the large thermal velocities of the neutrinos prevents their clustering on small scales. The bottom panels of that figure display the ratio between the matter power spectrum of the massive neutrinos to the massless neutrino model. We obtain the typical shape induced by massive neutrinos on that ratio, which can be explained by the extension of the halo model presented in \cite{Massara_2014}.

\begin{figure*}
\begin{center}
\includegraphics[width=0.95\textwidth]{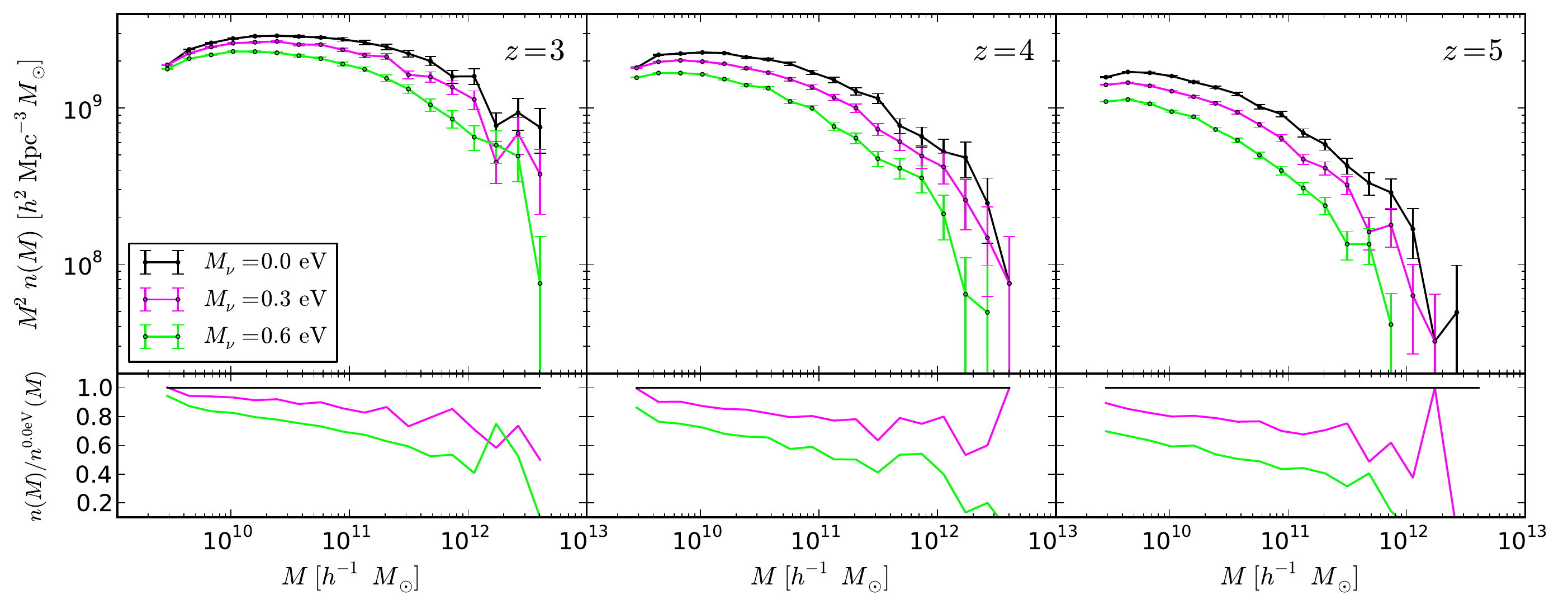}
\caption{Impact of massive neutrinos on the halo mass function obtained from our high-resolution hydrodynamic simulations. We show halo mass function for the model with massless neutrinos (black) and for the models with $M_\nu=0.3$ eV (magenta) and $M_\nu=0.6$ eV (green) neutrinos at $z=3$ (left), $z=4$ (middle) and $z=5$ (right). The error bars represent the uncertainty in the halo mass function assuming a Poisson distribution.}
\label{fig:Matter_clustering}
\end{center}
\end{figure*}

In Fig.~\ref{fig:Matter_clustering} we show the halo mass function of the cosmological models with massless and massive neutrinos at redshifts $z=3$, $z=4$, and $z=5$. We find that massive neutrinos suppress the abundance of dark matter halos, with the suppression increasing with the sum of the neutrino masses, with the halo mass, and with redshift. The weaker clustering of matter in cosmologies with massive neutrinos, with respect to their massless neutrino counterpart, is also what induces the suppression in the abundance of dark matter halos.

\end{document}